\documentclass[twocolumn,twocolappendix]{aastex63}
\pdfoutput=1 
\usepackage[T1]{fontenc}
\usepackage{ae,aecompl}
\usepackage[utf8]{inputenc}
\usepackage[english]{babel}
\usepackage{amsmath,amstext}
\usepackage{apjfonts}
\usepackage{wasysym}
\usepackage{chngcntr}
\usepackage[figure,figure*]{hypcap}
\usepackage[hang]{footmisc}
\shorttitle{Applying \textsc{UniverseMachine} to Zoom-in Simulations}
\shortauthors{Y.~Wang et al.}
\graphicspath{{./}{figures/}}

\begin{document}

\title{\textsc{UniverseMachine}: Predicting Galaxy Star Formation over Seven Decades of Halo Mass with Zoom-in Simulations} 

\correspondingauthor{Yunchong Wang}
\email{ycwang19@stanford.edu}

\author[0000-0001-8913-626X]{Yunchong Wang}
\affiliation{Department of Physics, Stanford University, 382 Via Pueblo Mall, Stanford, CA 94305, USA}
\affiliation{Kavli Institute for Particle Astrophysics \& Cosmology, P. O. Box 2450, Stanford University, Stanford, CA 94305, USA}

\author[0000-0002-1182-3825]{Ethan O.~Nadler}
\affiliation{Department of Physics, Stanford University, 382 Via Pueblo Mall, Stanford, CA 94305, USA}
\affiliation{Kavli Institute for Particle Astrophysics \& Cosmology, P. O. Box 2450, Stanford University, Stanford, CA 94305, USA}

\author[0000-0002-1200-0820]{Yao-Yuan Mao}
\altaffiliation{NASA Einstein Fellow}
\affiliation{Department of Physics and Astronomy, Rutgers, The State University of New Jersey, Piscataway, NJ 08854, USA}

\author[0000-0002-0298-4432]{Susmita Adhikari}
\affiliation{Kavli Institute for Particle Astrophysics \& Cosmology, P. O. Box 2450, Stanford University, Stanford, CA 94305, USA}
\affiliation{Department of Astronomy and Astrophysics, University of Chicago, Chicago, IL 60637, USA}
\affiliation{Kavli Institute for Cosmological Physics, University of Chicago, Chicago, IL 60637, USA}

\author[0000-0003-2229-011X]{Risa H.~Wechsler}
\affiliation{Department of Physics, Stanford University, 382 Via Pueblo Mall, Stanford, CA 94305, USA}
\affiliation{Kavli Institute for Particle Astrophysics \& Cosmology, P. O. Box 2450, Stanford University, Stanford, CA 94305, USA}
\affiliation{SLAC National Accelerator Laboratory, Menlo Park, CA 94025, USA}

\author[0000-0002-2517-6446]{Peter Behroozi}
\affiliation{Steward Observatory and Department of Astronomy, University of Arizona, Tuscon, AZ 85721}

\begin{abstract}

We apply the empirical galaxy--halo connection model \textsc{UniverseMachine} to dark matter-only zoom-in simulations of isolated Milky Way (MW)--mass halos along with their parent cosmological simulations. This application extends \textsc{UniverseMachine} predictions into the ultra-faint dwarf galaxy regime ($ 10^{2}\,\mathrm{M_{\astrosun}} \leqslant M_{\ast} \leqslant 10^{5}\,\mathrm{M_{\astrosun}}$) and yields a well-resolved stellar mass--halo mass (SMHM) relation over the peak halo mass range $10^8\,\mathrm{M_{\astrosun}}$ to $10^{15}\,\mathrm{M_{\astrosun}}$. The extensive dynamic range provided by the zoom-in simulations allows us to assess specific aspects of dwarf galaxy evolution predicted by \textsc{UniverseMachine}. In particular, although \textsc{UniverseMachine} is not constrained for dwarf galaxies with $M_* \lesssim 10^{8}\,\mathrm{M_{\astrosun}}$, our predicted SMHM relation is consistent with that inferred for MW satellite galaxies at $z=0$ using abundance matching. However, \textsc{UniverseMachine} predicts that nearly all galaxies are actively star forming below $M_{\ast}\sim 10^{7}\,\mathrm{M_{\astrosun}}$ and that these systems typically form more than half of their stars at $z\lesssim 4$, which is discrepant with the star formation histories of Local Group dwarf galaxies that favor early quenching. This indicates that the current \textsc{UniverseMachine} model does not fully capture galaxy quenching physics at the low-mass end. We highlight specific improvements necessary to incorporate environmental and reionization-driven quenching for dwarf galaxies, and provide a new tool to connect dark matter accretion to star formation over the full dynamic range that hosts galaxies.

\end{abstract}

\keywords{\href{http://astrothesaurus.org/uat/353}{Dark matter (353)}; \href{http://astrothesaurus.org/uat/416}{Dwarf galaxies (416)}; \href{http://astrothesaurus.org/uat/594}{Galaxy evolution (594)}; \href{http://astrothesaurus.org/uat/1965}{Computational methods (1965)}}

\section{Introduction} 
\label{sec:1}

Hierarchical structure formation via cold-dark-matter (CDM) collapse in a flat-$\Lambda$CDM cosmology model~\citep{1978MNRAS.183..341W,1984Natur.311..517B,1991MNRAS.251..128L,1994MNRAS.271..781C} has become the standard paradigm for galaxy formation over the past few decades. Dwarf galaxies, which are tracers of the small-scale structures of $\Lambda$CDM (with peak halo mass $M_{\mathrm{Peak}} \lesssim 10^{11}\,\mathrm{M_{\astrosun}}$), have shown diverse and initially surprising features in observations compared to the predictions from cold-dark-matter (CDM) theory and numerical simulations (see \citealt{2017ARA&A..55..343B} for a review). It is crucial to understand the detailed formation mechanisms of dwarf galaxies, as well as the evolution of dwarf galaxies and the (sub)halos in which they form. This understanding could have a profound impact on our ability to test galaxy formation and the flat-$\Lambda$CDM paradigm by fully harnessing observations across the full range of galaxy mass scales. Key features of dwarf galaxies such as star formation histories~\citep{1998ARA&A..36..435M,2005MNRAS.362...41G,2009ARA&A..47..371T,2012ApJ...757...85G,2014ApJ...789..147W,2014ApJ...789..148W,2015ApJ...804..136W} and quenched (not-star-forming) fractions as a function of stellar mass~\citep{2012AJ....144....4M,2013AJ....146...46K,2016MNRAS.463.1916F,2021ApJ...907...85M} have been studied extensively in observations.

To connect the observed properties of galaxies to the dark matter (sub)halos in which they reside, two approaches of modeling the galaxy--halo connection have been widely adopted: (\emph{i}) models that are motivated by the underlying physics related to galaxy formation and gravity governing both dark matter and baryons (gas and stars), such as hydrodynamic simulations (see \citealt{2020NatRP...2...42V} for a review) and semi-analytic models (SAMs; see  \citealt{2015MNRAS.451.2663H,2020arXiv201104670Y,2020arXiv200505974J} for some recent examples), or (\emph{ii}) models that are motivated by empirical relations between galaxies and dark matter halos such as subhalo abundance matching and halo occupation distribution methods (see \citealt{2015ARA&A..53...51S,2018ARA&A..56..435W} for reviews). Although these classes of models have each been effective for studying many aspects of galaxy--halo connection modeling, they possess different limitations when applied to the low-mass end of the observable galaxy population. The former approach relies on the assumed underlying physical subgrid or semi-analytic models for baryons, that require high resolution (and thus demanding computational resources to resolve dwarf galaxies in cosmological volumes), and that are less easily adaptable to be improved by observational constraints. The latter approach is more flexible due to its statistical description and thus can more easily adapt to new measurements, but makes strong assumptions on the number and density of central and satellite galaxies in connection to their host (sub)halos. All approaches can require calibration of systematics such as orphan galaxies ~\citep[e.g.,][]{2005Natur.435..629S,2017MNRAS.469..749P,2019ApJ...873...34N}, although the numerical aspects of this are different for the different modeling types.

The empirical forward modeling approach (e.g.,, \citealt{2013ApJ...770...57B,2013MNRAS.428.3121M,2017MNRAS.470..651R,2018MNRAS.477.1822M,2019MNRAS.488.3143B}) has inherited some merits from both branches above (we refer to this approach as an `empirical model' in the rest of the paper). Current empirical models paint the star formation rates (SFRs) of galaxies onto dark matter halos in $N$-body simulations using flexible mathematical formulae. These have the advantage of requiring fewer assumptions about the underlying galaxy formation processes than do hydrodynamic simulations and semi-analytic models. They are also more flexible in connecting galaxy and halo properties than standard abundance matching and halo occupation distribution models. 

In this work, we use the empirical model \textsc{UniverseMachine}~(\citealt{2019MNRAS.488.3143B}, B19 hereafter) to study the galaxy--halo connection in the low-mass dwarf galaxy regime. The model parameterizes the distribution of galaxies' SFRs as a function of halo mass (maximum circular velocity $v_{\mathrm{max}}$), halo accretion rate (change in $v_{\mathrm{max}}$ over time), and redshift. The SFRs of galaxies are drawn randomly from these distributions and then painted onto halos. This process is conducted self-consistently to match a comprehensive set of observational constraints at the intermediate- to high-mass end: stellar mass functions (stellar mass $M_{\ast}/\,\mathrm{M_{\astrosun}}\in[10^{7}, 10^{12}]$, redshift $z\in[0, 8]$), galaxy quenched fractions ($M_{\ast}\gtrsim 10^{9}\,\mathrm{M_{\astrosun}}$, $z\lesssim 4$), the cosmic star formation history ($z\leqslant 10$), specific SFRs (SFR per unit stellar mass, $M_{\ast}\gtrsim 10^{8}\,\mathrm{M_{\astrosun}}$, $z\leqslant 8$), UV luminosity functions ($M_{\ast}\gtrsim 10^{8}\,\mathrm{M_{\astrosun}}$, $z\in[4, 8]$), and galaxy clustering information from auto- and cross- correlations of different galaxy populations ($M_{\ast}/\,\mathrm{M_{\astrosun}}\in[10^{10}, 10^{12}]$, redshift $z\lesssim 1$) (see Section 3 and Appendix C in B19 for more details), among others. The model also accounts for short-timescale SFR variations, stellar population mass loss, galaxy mergers, evolution of the intra-cluster-light (ICL) stellar population, orphan galaxies due to subhalo disruption, dust and metallicity, and observational systematics.

The \textsc{UniverseMachine} Data Release 1 (UM DR1) was calibrated using the cosmological simulation {\it Bolshoi-Planck}~\citep{2016MNRAS.457.4340K,2016MNRAS.462..893R} with the above-mentioned set of observational data with stellar masses $M_{\ast}\gtrsim 10^{7}\,\mathrm{M_{\astrosun}}$ (corresponding to peak halo masses $M_{\mathrm{Peak}}\gtrsim 10^{10}\,\mathrm{M_{\astrosun}}$). The dark matter resolution of $\sim10^{8}\,\mathrm{M_{\astrosun}}$ for the {\it Bolshoi--Planck} simulation is not ideal for studying the dwarf galaxy regime ($M_{\ast}\leqslant 10^{9}\,\mathrm{M_{\astrosun}}$ and $M_{\mathrm{Peak}}\lesssim 10^{11}\,\mathrm{M_{\astrosun}}$). The \textsc{UniverseMachine} can be run on higher-resolution large-volume cosmological simulations, as was done in \cite{2020MNRAS.499.5702B} to predict observables for the JWST, but this approach cannot gain much resolution without an equivalent sacrifice in volume and sample statistics. 

We take an exploratory approach in this work, that extrapolates the \textsc{UniverseMachine} model into the dwarf galaxy regime by applying it to zoom-in simulations of MW-sized halos ($M_{\mathrm{vir}}\approx 10^{12} \,\mathrm{M_{\astrosun}}$). These simulations resolve subhalos down to $\sim 10^{8}\,\mathrm{M_{\astrosun}}$, corresponding to the halos that host the smallest known galaxies, ultra-faint dwarf galaxies with stellar masses as low as $M_{\ast}\sim 10^{2}\,\mathrm{M_{\astrosun}}$. By qualitatively comparing the extrapolated \textsc{UniverseMachine} predictions to observational constraints, we can identify specific limitations---especially pertaining to quenching physics---that the DR1 model failed to capture. Thus, we provide a framework for exploring the impact of reionization and environmental quenching mechanisms on dwarf galaxies in the context of a dark matter accretion-based model for galaxy assembly.

Since \textsc{UniverseMachine} is currently constrained only at the intermediate- to high-mass end, quenching mechanisms for dwarf galaxies are not explicitly included in the DR1 version of the model. In particular, the model assumption that correlates galaxy star formation rate with halo accretion rate may break down at the epoch of reionization; in this epoch, photo-ionization in dwarfs with $M_{\ast}\lesssim 10^{5}\,\mathrm{M_{\astrosun}}$ likely prevents gas from cooling~\citep{1996ApJ...465..608T,2000ApJ...539..517B,2002MNRAS.333..156B,2012ApJ...759L..38A,2014ApJ...796...91B}. It is also important to investigate if the current modeling of environmental quenching primarily through parameterized galaxy quenched fractions constrained at high stellar masses is sufficient to capture galaxy evolution in the dwarf regime. Recent and ongoing observations of dwarf galaxies in the low-redshift Universe (e.g.,\ from the SAGA Survey, ~\citealt{2017ApJ...847....4G,2021ApJ...907...85M}, and satellites in the Local Volume~\citealt{2021ApJ...908..109C,2020ApJ...891..144C,2021MNRAS.500.3854D}) as well as near the Milky Way (MW) (e.g.,\ from the Dark Energy Survey~\citealt{2015ApJ...813..109D,2020ApJ...893...47D,2020ApJ...893...48N}), along with upcoming observations that will extend such observations in both the local and distant universe (e.g.,\ from campaigns including the Vera C.\ Rubin Observatory, JWST, and the Nancy Grace Roman Space Telescope), promise to deliver tighter observational constraints on the model assumptions regarding dwarf galaxy evolution.

This paper is organized as follows: in Section~\ref{sec:2}, we introduce the dark matter simulations underlying this work. We describe our method for applying \textsc{UniverseMachine} to zoom-in and cosmological simulations in Section \ref{sec:3}, and describe the halo mass functions, stellar mass--halo mass relations, and matched evolution histories of systems in both resolution levels. In Section~\ref{sec:4}, we study the low-mass end of the stellar mass--halo mass relation, its redshift evolution, the star formation histories matched across our simulation resolution levels, and the predicted quenched fractions for central and satellite galaxies. We discuss future modeling directions in Section~\ref{sec:5} and we summarize our main findings and discuss the outlook for empirically modeling the low-mass end of the galaxy--halo connection in Section \ref{sec:6}.

\section{Simulations}
\label{sec:2}

\subsection{The Chinchilla suite}
\label{sec:2.1}

We begin by introducing the dark matter-only simulations on which we base our \textsc{UniverseMachine} analyses. The Chinchilla simulation suite is a set of dark-matter-only $N$-body simulations run with the \textsc{L-Gadget} code (an $N$-body only variation of the \textsc{Gadget-2} code, \citealt{2005MNRAS.364.1105S}) under a flat $\Lambda$CDM cosmology with parameters $H_{0} = 70\ \mathrm{km\,s^{-1}\,Mpc^{-1}}$ ($h=0.7$), $\Omega_{\mathrm{m}} = 0.286$, $\Omega_{\Lambda} = 0.714$, $n_{\mathrm{s}} = 0.96$, and $\sigma_{8} = 0.82$. Dark matter halos are identified using the phase-space halo finder \textsc{Rockstar}~\citep{2013ApJ...762..109B} with a virial overdensity threshold of $\Delta_{c, \mathrm{vir}}\approx 99.2$ at $z=0$, while merger trees are constructed using \textsc{Consistent-Trees}~\citep{2013ApJ...763...18B}. Halos contained within the virial radius of a more massive nearby halo are defined as subhalos, other halos are defined as host halos. 

Here we use only the two smallest cosmological boxes in the suite, first introduced in \citealt{2015ApJ...810...21M}. 
These boxes both have a side length of 125 $\mathrm{Mpc}\ h^{-1}$ with periodic boundary conditions. The higher resolution box has $2048^{3}$ resolution elements with dark matter particle mass of $m_{\mathrm{DM}} = 1.80\times 10^{7}\,\mathrm{M_{\astrosun}}\,h^{-1}$ (denoted as \textsc{c125-2048} hereafter), and a softening length of $0.5\  \mathrm{kpc}\,h^{-1}$ (Plummer-equivalent force softening scale). The lower resolution box has $1024^{3}$ resolution elements with particle masses of $m_{\mathrm{DM}} = 1.44\times 10^{8}\,\mathrm{M_{\astrosun}}\,h^{-1}$ (denoted as \textsc{c125-1024} hereafter), and a softening length of $1.0\ \mathrm{kpc}\,h^{-1}$. The boxes are initialized with the same set of initial conditions generated using \textsc{2LPTic}~\citep{2006MNRAS.373..369C} implementing the matter power spectrum from \textsc{camb}\footnote{\url{https://camb.info/}}, although \textsc{c125-2048} starts at $z=199$, while \textsc{c125-1024} starts at $z=99$. Their halo catalogs and merger trees both start from a scale factor of $a=0.075$ ($z\approx 12.3$) with 100 snapshots down to $a=1$ ($z=0$).

\subsection{The MW zoom-ins}
\label{sec:2.2}

The 45 MW-size halos chosen for the zoom-in re-simulations were isolated hosts selected from the parent box \textsc{c125-1024}, with host halo masses in the range of $ M_{\mathrm{vir}}/\,\mathrm{M_{\astrosun}}\in 10^{12.1 \pm 0.03}$~\citep{2015ApJ...810...21M}. This mass range is consistent with the latest halo mass constraints of the MW inferred from the proper motion of Globular Clusters in \emph{ Gaia} DR2~($1.3\pm 0.3\times 10^{12}\,\mathrm{M_{\astrosun}}$ from \citealt{2019A&A...621A..56P}, $1.54^{+0.75}_{-0.44}\times 10^{12}\,\mathrm{M_{\astrosun}}$ from \citealt{2019ApJ...873..118W}). The initial conditions for each individual zoom-in halo are then generated with \textsc{Music}~\citep{2011MNRAS.415.2101H} to achieve the equivalent resolution of $8192^{3}$ elements in an initial (at $z=99$) Lagrangian region of $\sim 10 R_{\mathrm{vir}}$ centered on the MW host, while matching to the parent box on large scales. The high resolution particles in the zoom-in region have masses of $m_{\mathrm{DM}} = 2.82\times 10^{5}\,\mathrm{M_{\astrosun}}\,h^{-1}$ and softening lengths of $170\ \mathrm{pc}\,h^{-1}$. 

The halo catalogs and merger trees for the zoom-ins begin at $a=0.05$ ($z=19$) with 236 snapshots down to $a=1$ ($z=0$). At $z=0$, roughly $90\%$ of the total mass within $r_{90} \approx 3\ \mathrm{Mpc}$ 
of each MW host halo consists of high-resolution particles. Thus, the zoom-in simulations provide well-resolved subhalos in their infalling, splashback, accreted, and disrupted evolution phases far beyond the MW virial radius ($\sim 300$ kpc).
We summarize the key properties of our 45 MW zoom-in simulations and their two parent boxes in Table~\ref{tab:1}.

\begin{table*}
		\begin{center}
		\begin{tabular}{lccccc}
			\hline
			\hline
			Simulation & Box size [$\mathrm{Mpc}\,h^{-1}$] & Resolution elements & Particle mass [$\mathrm{M_{\astrosun}}\,h^{-1}$] & Softening scale [$\mathrm{kpc}\,h^{-1}$]\\
			\hline
			Zoom-ins  & - & $8192^{3}$ ($\mathrm{equivalent}$) & $2.82\times10^{5}$ & 0.17\\
			c125-2048 & 125 & $2048^{3}$ & $1.80\times 10^{7}$ & 0.50 \\
			c125-1024 & 125 & $1024^{3}$ & $1.44\times 10^{8}$ & 1.00 \\
			\hline
		\end{tabular}
        \end{center}
		\caption{The basic information for our 45 MW zoom-in simulations and their two parent boxes.}
		\label{tab:1}
\end{table*}

\section{Methods and Validation}
\label{sec:3}

We now overview the fundamental aspects of the empirical galaxy--halo connection model \textsc{UniverseMachine} (Section \ref{sec:3.1}). We present the halo mass function and the \textsc{UniverseMachine}-predicted stellar mass--halo mass (SMHM) relation across the three simulation resolutions in Section \ref{sec:2}, highlighting our procedure for applying \textsc{UniverseMachine} to zoom-in simulations (Section \ref{sec:3.2}). We study the dark matter halo evolution and predicted star formation histories of matched objects in the zoom-in re-simulations and their parent boxes (Section \ref{sec:3.3}).

\subsection{\textnormal{\textsc{UniverseMachine}} Model}
\label{sec:3.1}

\textsc{UniverseMachine}\footnote{\url{https://bitbucket.org/pbehroozi/universemachine}}~(B19) is an empirical galaxy--halo connection model that self-consistently predicts the star formation rates (SFRs) of galaxies inside dark matter halos given the halo mass (parameterized by the maximum circular velocity $v_{\mathrm{max}}$), assembly history (parameterized by halo formation time and dark matter accretion rate), and redshift. It is empirical in the sense that the probability distribution of galaxy SFR as a function of halo properties and evolution is informed by observational constraints as opposed to assumptions about galaxy physics \citep{2018ARA&A..56..435W}.

The model starts by choosing a point from the parameter space of redshift-dependent analytic relations between SFR and the quenched galaxy fraction $f_{\mathrm{Quench}}$ versus halo $v_{\mathrm{max}}$. This leads to a probability distribution of SFR at each set of fixed $v_{\mathrm{max}}$ and redshift values. The model then integrates the assigned SFRs of the different halos from high redshift down to $z=0$ along the halo merger trees to obtain the full stellar mass evolution history. Afterwards, intrinsic observables of the baryonic properties corresponding to the halo population including the stellar mass functions, galaxy quenched fractions, cosmic star formation history, and galaxy auto- and cross- correlations as a function of redshift are constructed from the SFR and stellar mass histories. Corrections to these intrinsic observables are applied to mimic observational biases and uncertainties. The corrected observable values are compared to observations to obtain a likelihood for the parameters governing the analytic SFR($v_{\mathrm{max}}$) and $f_{\mathrm{Quench}}(v_{\mathrm{max}})$ relations, which are then sampled in a Monte Carlo Markov Chain (MCMC) framework to derive the posterior distribution over the \textsc{UniverseMachine} model parameters.

Here we implement the `best-fit' UM DR1 model from B19, using the set of 44 parameters that maximize the likelihood including the SFR($v_{\mathrm{max}}$) and $f_{\mathrm{Quench}} (v_{\mathrm{max}})$ relations, cosmic star formation history, specific star formation rate as a function of stellar mass, galaxy auto- and cross- correlations  given the constraints of observed SFR distributions in halos. The assumed cosmic baryon fraction is $f_{\mathrm{b}}=0.158$. We emphasize that the UM DR1 model was calibrated using the {\it Bolshoi-Planck} simulation~\citep{2016MNRAS.457.4340K,2016MNRAS.462..893R} which has a particle resolution of $m_{\mathrm{DM}}\sim 2.2 \times 10^{8} \,\mathrm{M_{\astrosun}}$ (box size of 250 Mpc$\mathrm{h}^{-1}$ and $2048^{3}$ particles). Thus, the majority of the predicted stellar mass function and quenched fraction constraints were derived in the stellar mass range of $M_{\ast} \gtrsim 10^{8}\,\mathrm{M_{\astrosun}}$ (corresponding to a peak halo mass range of $M_{\mathrm{Peak}}\sim 6\times 10^{10}\,\mathrm{M_{\astrosun}}$), while the galaxy auto- and cross- correlation constraints were largely in the $10^{10}\,\mathrm{M_{\astrosun}}\lesssim M_{\ast} \lesssim 10^{12}\,\mathrm{M_{\astrosun}}$ range (corresponding to $M_{\mathrm{Peak}}/\,\mathrm{M_{\astrosun}} \in [10^{11.8}, 10^{15}]$). 

A major aim of this work is therefore to apply the UM DR1 model to zoom-in re-simulations in a mass range that was unresolved and unconstrained in B19. We refer the reader to Section 3 and Appendix C of B19 for a detailed description of the \textsc{UniverseMachine} model, observational constraints, systematics, and uncertainties. In the following, unless stated explicitly to be the intrinsic values (e.g.,\ in Fig.~\ref{fig:3} and Fig.~\ref{fig:5}), we use the {\it corrected} observable values of the \textsc{UniverseMachine} output SFR and $M_{\ast}$. Because the parent boxes and the zoom-ins cover a vast dynamic range in mass resolution, we set the $v_{\mathrm{max}}$ thresholds for orphans (subhalos above which are resolved and traced for their orphan galaxy after disruption)~\footnote{The fiducial \textsc{UniverseMachine} orphan model and mass-loss prescription is described in B19 and is similar to the model in~\cite{2016MNRAS.458.2848J}.} to $10\ \mathrm{km\,s^{-1}}$, $35\ \mathrm{km\,s^{-1}}$, and $60\ \mathrm{km\,s^{-1}}$ for the joint 45 zoom-ins, \textsc{c125-2048}, and \textsc{c125-1024}, respectively, following the resolution cuts defined in Appendix~\ref{sec:A2}.

\subsection{Applying \textsc{UniverseMachine} to Zoom-ins and Chinchilla}
\label{sec:3.2}

\begin{figure*}[ht]
    \centering
	\includegraphics[width=2\columnwidth]{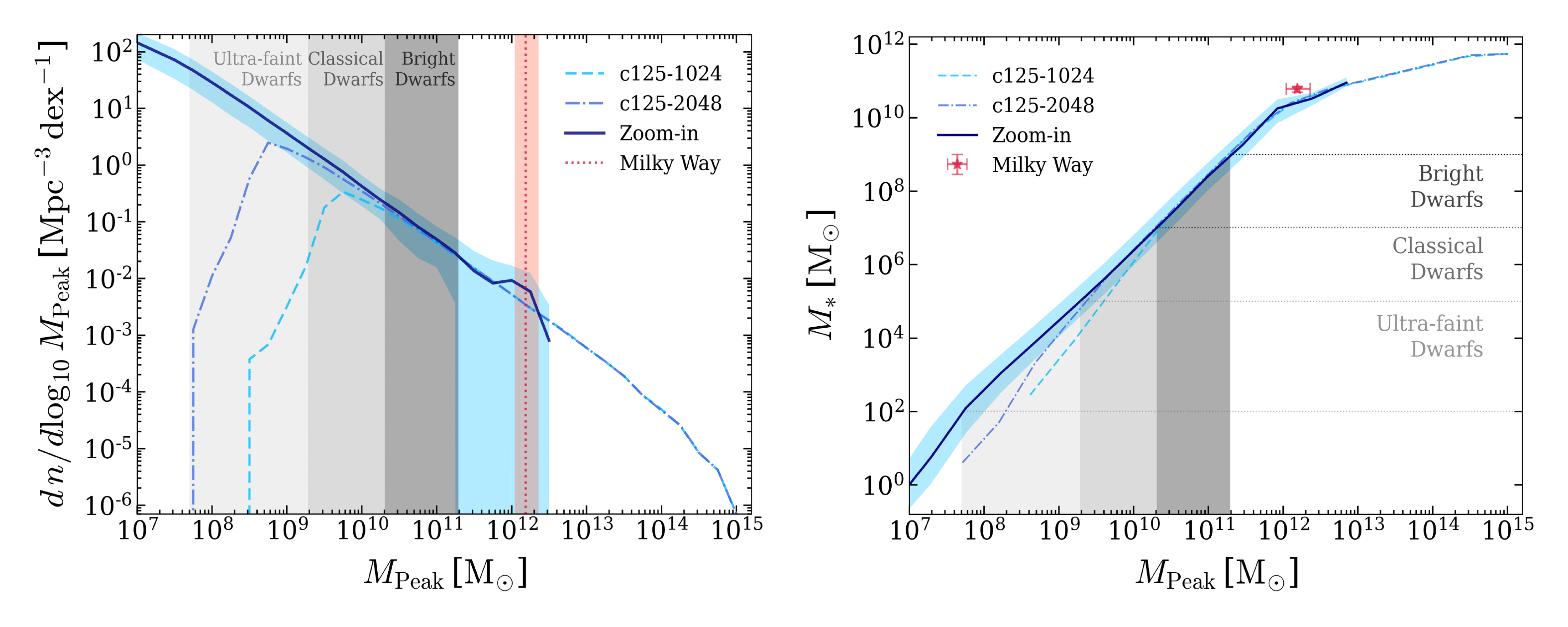}
    \caption{{\it Left panel:} Peak halo mass function of the three different resolution simulations at $z=0$ used in this work. The solid, dot-dashed, and dashed curves respectively indicate halo mass functions for our 45 MW zoom-in re-simulations and the \textsc{c125-2048} and \textsc{c125-1024} boxes from which these halos are drawn. The shaded region represents the standard deviation due to host-to-host scatter for the 45 MW zoom-ins. The bump in the MW re-simulations curve at $\sim10^{12}\,\mathrm{M_{\astrosun}}$ results from selecting for MW-mass host halos in the zoom-in regions. {\it Right panel:} Median stellar mass--halo mass relations ($z=0$) obtained by applying the UM DR1 model to each resolution level, shown with the same color scheme and line styles as the left panel. The blue band shows 68\% scatter of the SMHM relation for the 45 MW zoom-ins; the two cosmological boxes have almost identical levels of scatter. The ``bright,'' ``classical,'' and ``ultra-faint'' dwarf galaxy regimes defined in \citet{2017ARA&A..55..343B} are illustrated as gray bands and lines. Observational estimates of the MW halo mass~\citep{2019ApJ...873..118W} and stellar mass~\citep{2015ApJ...806...96L} are shown by the red band and error bar, respectively.}
    \label{fig:1}
\end{figure*}

We present the halo mass function and the SMHM relation derived from applying the UM DR1 model to 45 MW zoom-ins and their Chinchilla parent boxes. We also discuss challenges involved in applying \textsc{UniverseMachine} to zoom-in simulations and the steps taken to overcome them. 

The left panel of Fig.~\ref{fig:1} shows the halo mass functions of the three simulation resolutions.  We follow previous works (e.g.,  \citealt{2013ApJ...770...57B,2014MNRAS.438.2578G}) by using peak halo mass, the maximum value of the halo virial mass along the main progenitor branch in its merger tree, for all halo masses. The peak halo mass does not depend on subhalo mass loss due to tidal stripping after infall and correlates better with stellar mass than present-day virial mass (e.g., \citealt{2013ApJ...770...57B,2013MNRAS.428.3121M, 2013ApJ...771...30R}). The joint zoom-in halo number density $n$ is calculated as the average number density in the 45 zoom-in regions set by the zoom-in region volume factors in the initial Lagrangian spaces. We note that our 45 zoom-in MW halos are isolated and they tend to live in slightly underdense environments, with the average matter density (measured within a sphere with radius of $20\ \mathrm{Mpc}/h$ centered on the host halo) $\sim0.87\times$ the mean matter density of the parent cosmological boxes.

The three resolution boxes show overall consistent halo mass functions in the peak halo mass range of $M_{\mathrm{Peak}}/\mathrm{M_{\astrosun}}\in$[$10^{10}$, $10^{11}$], and the two cosmological boxes extend all the way up to the cluster mass scale of $\sim 10^{15}\,\mathrm{M_{\astrosun}}$ (with their Lagrangian volumes same as their box sizes). At the low-mass end, the joint mass function of the 45 MW zoom-in re-simulations extends the dynamic range of our joint halo mass function far below $\sim 10^{9}\,\mathrm{M_{\astrosun}}$, where both cosmological boxes are poorly resolved. The zoom-ins cut off right above the MW halo mass of a few times $10^{12}\,\mathrm{M_{\astrosun}}$, and the hump at that mass reflects the selection of MW-sized host halos in the zoom-in regions. An estimate of the MW halo mass derived from \emph{Gaia} data \citep{2019ApJ...873..118W}, which is consistent with other recent MW halo mass constraints, is shown for comparison.

In the right panel of Fig.~\ref{fig:1}, we show the SMHM relations for the three simulation resolutions with the UM DR1 model applied to each box. The median SMHM relations coincide well (for both the normalization and slope) in the halo mass range $M_{\mathrm{Peak}}/\mathrm{M_{\astrosun}}\in$[$10^{10}$, $10^{11}$], where the halo mass functions are also converged. The most striking feature of the SMHM relation at the low-mass end is that all three resolutions follow a power-law relation with the same slope that extends from the knee of the relation at around $10^{12}\,\mathrm{M_{\astrosun}}$ down to the resolution limit of each simulation. This suggests that the SMHM relation in the zoom-ins is an extrapolation of the SMHM relation in the UM DR1 model, which is constrained down to $M_{\mathrm{Peak}}\sim 10^{11}\,\mathrm{M_{\astrosun}}$. We discuss the implications of our predicted SMHM relation in Section~\ref{sec:4}. We also quote and show the observational value of the MW stellar mass~\citep{2015ApJ...806...96L} along with the three dwarf galaxy regimes as defined in \citet{2017ARA&A..55..343B} for comparison. This work defines bright dwarfs as galaxies with stellar mass in the range of $M_{\ast}/\mathrm{M_{\astrosun}}\in[10^{7}, 10^{9}]$, classical dwarfs with stellar masses of $M_{\ast}/\mathrm{M_{\astrosun}}\in[10^{5}, 10^{7}]$ , and ultra-faint dwarfs with stellar masses of $M_{\ast}/\mathrm{M_{\astrosun}}\in[10^{2}, 10^{5}]$ . We convert the stellar masses in these three regimes back to the peak halo mass intervals using the median SMHM relation in the joint zoom-ins, and show them as the shaded regions in the left panel of Fig.~\ref{fig:1}. The MW zoom-ins provide sufficient resolution to model the smallest-known galaxies, down to the lower mass limit of the ultra-faint dwarfs (UFDs), a regime that is crucial for studies of galaxy formation~\citep{2010AdAst2010E..21W,2012AJ....144....4M,2015ApJ...807...50B} and dark matter microphysics~\citep{2020arXiv200800022N}.

Applying the \textsc{UniverseMachine} model to zoom-in simulations is non-trivial for several technical reasons. In particular, individual zoom-in realizations of single host halos lead to poor sampling of the halo mass function at halo masses around the MW host, and also provide only a few realizations of the most massive subhalos. This renders the \textsc{UniverseMachine}'s ranking algorithm, which assigns higher SFR to halos with higher accretion rates ($\Delta v_{\mathrm{max}}$, the change in $v_{\mathrm{max}}$ over one dynamical timescale) based on the full span of accretion rates at fixed halo mass, more susceptible to stochasticity, leading to biased rank predictions. We illustrate in Appendix~\ref{sec:A1} how we overcome this by joining 45 MW zoom-ins together to achieve a halo distribution that is largely consistent with the parent cosmological boxes. In summary, we implement two important changes when applying \textsc{UniverseMachine} to zoom-in simulations as opposed to cosmological boxes\footnote{Detailed steps can be found in the `Running on Different Simulations' section on: \url{https://bitbucket.org/pbehroozi/universemachine/src/main/}. The two changes mentioned in the text are made in step 1 and step 4.}:
\begin{enumerate}
    \item We join the \textsc{Rockstar} halo catalogs together for a large sample of zoom-in boxes at every simulation snapshot, such that each distinct (sub)halo is included exactly once in the joint halo catalog. In our case, this is equivalent to packing all 45 MW zoom-in halos together into one cosmological box with the same size as their parent boxes.
    \item \textsc{UniverseMachine} performs halo ranking with a spline interpolation step. We carry out this step with the joint halo catalog. Ranks produced in this manner will be more accurate due to the more comprehensive sampling of the halo mass function, especially at the high-mass end.
\end{enumerate}

In this way, we join a large number of zoom-in simulations to self-consistently extend \textsc{UniverseMachine} predictions into the ultra-faint dwarf galaxy regime. We note that the rank distributions, described in Appendix \ref{sec:A1} and shown in Fig.~\ref{fig:A1}, are approximately self-similar when expressed in terms of particle number. Thus, step 1 could potentially be circumvented by scaling a given rank distribution to match those shown in Fig.~\ref{fig:A1}. This analytic procedure would yield approximate, unbiased ranks for halos in higher-resolution simulations and is useful for simulation suites that do not include a substantial number of high-resolution simulations.

\subsection{Validation: Matched Subhalo Evolution}
\label{sec:3.3}

\begin{figure*}[t]
    \centering
	\includegraphics[width=2\columnwidth]{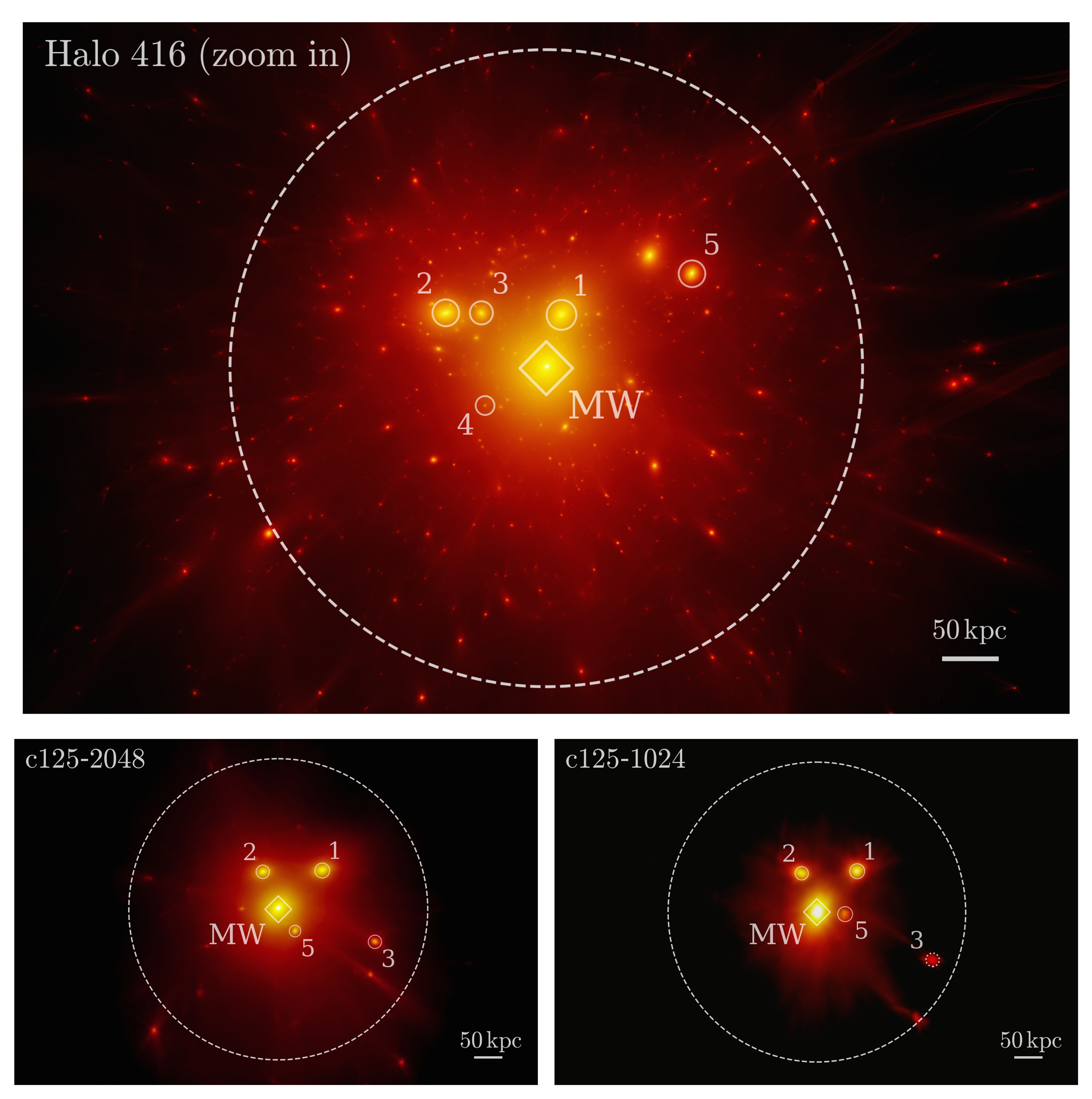}
    \caption{Projected dark matter density map of one of our MW-mass zoom-in simulations (top panel) and the matched system in multiple resolutions of its parent simulation (bottom panels). The images are centered on the MW host halo (diamonds) and viewed from an angular diameter distance of 1.75 $\mathrm{Mpc}\ h^{-1}$ with a vertical aperture of $15^{\circ}$. Five subhalos in the zoom-in simulation and their matched counterparts in the parent boxes are denoted by open circles. The dark matter growth and star formation histories of these systems are shown in Fig.~\ref{fig:3}. Subhalo 4 corresponds to an ultra-faint dwarf galaxy that only exists in the zoom-in simulation; similarly, Subhalo 3 is not well resolved in \textsc{c125-1024} and is therefore marked with a dotted circle. Subhalo 5 is a splashback system in the zoom-in simulation at $z=0$ (although it appears inside the MW virial radius in projection), while it is inside the MW virial radius in the parent boxes at the final snapshot. The apparent difference in the locations of subhalos in different resolutions are due to the difference in their orbital phases. The large dashed circle indicates the virial radius of the MW host ($\sim 300$ kpc).}
    \label{fig:2}
\end{figure*}

\begin{figure*}
    \centering
	\includegraphics[width=\textwidth]{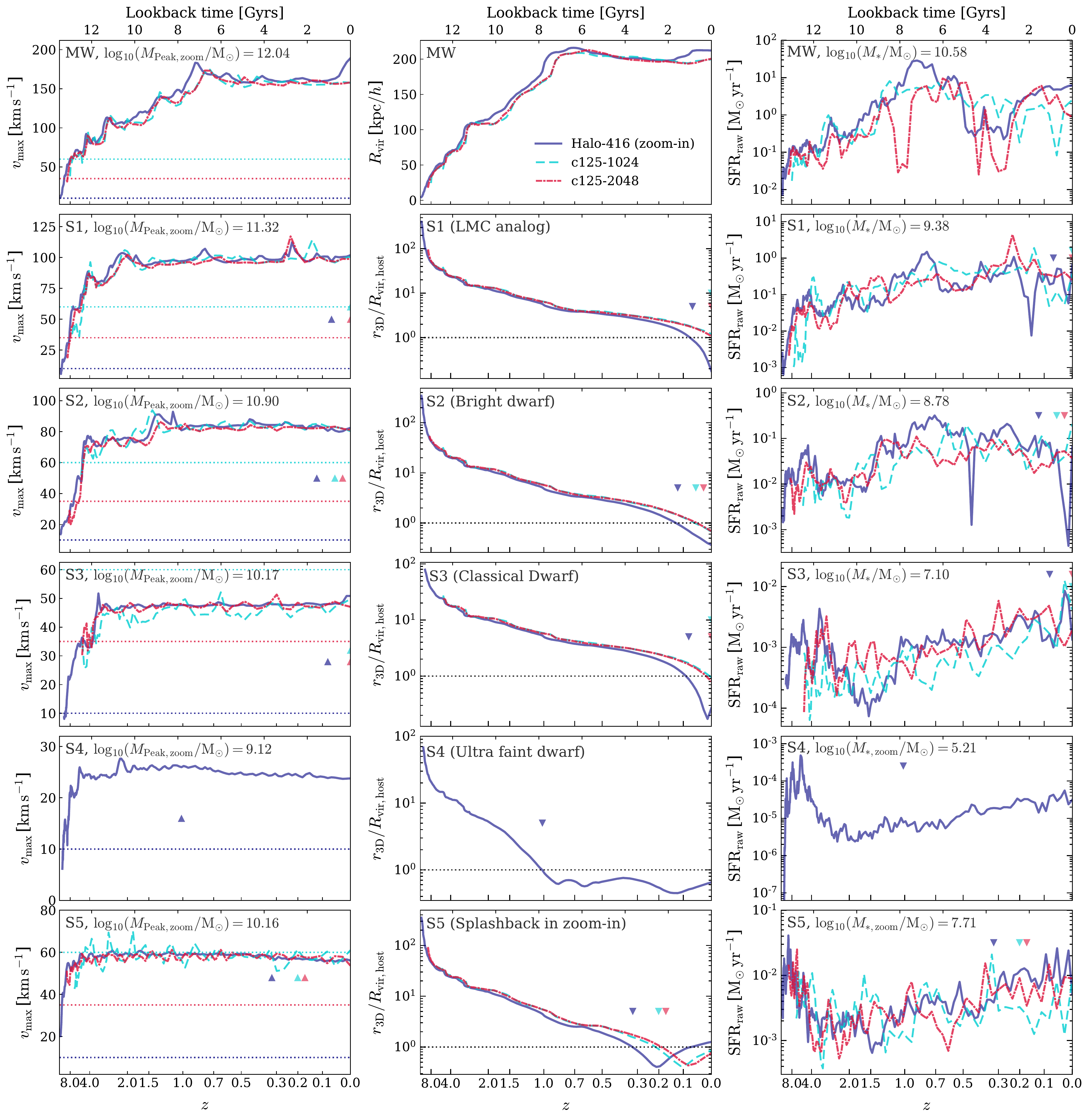}
    \caption{Evolution of one of our MW-mass zoom-in simulation host halos and five subhalos (labeled as S1 to S5) surrounding it at each of our resolution levels (these systems are illustrated in Fig.~\ref{fig:2}). The left column shows the $v_{\mathrm{max}}$ evolution, indicating the dark matter accretion history of these systems, the middle column shows the evolution of the MW host virial radius ($R_{\mathrm{vir}}$) and the 3D distance of the satellites $r_{\mathrm{3D}}$ to the host in units of $R_{\mathrm{vir, host}}$, and the right column shows the evolution of the intrinsic star formation rate predicted by \textsc{UniverseMachine} ($\mathrm{SFR_{raw}}$). Triangles with the same color as the evolution curves in the left, middle and right columns indicate the infall times of the subhalos (the first crossing of $R_{\mathrm{vir, host}}$ in the middle column). Horizontal dotted lines in the left column panels of the same color indicate resolution thresholds, defined in Appendix~\ref{sec:A2} for our three resolution levels. We label the peak halo masses in the left column and the `observed' stellar masses predicted by \textsc{UniverseMachine} at $z=0$ in the right column for the zoom-in (sub)halos. We note that the SFR evolution is not expected to be identical across different resolutions (even for a fixed $v_{\mathrm{max}}$ history) due to the different random numbers assigned to halos in each simulation.}
    \label{fig:3}
\end{figure*}

In the previous subsection, we showed that the halo mass function and the SMHM relation are consistent across the zoom-in and cosmological boxes. Since our three different resolution simulations share the same initial conditions, we now illustrate \textsc{UniverseMachine} predictions in detail by studying the dark matter growth and star formation histories of matched (sub)halos across the three boxes.

Fig.~\ref{fig:2} shows the projected dark matter density maps for the zoom-in MW halo No.\ 416 and its matched counterparts in the two parent boxes.\footnote{This is one of the MW-like halos that includes a realistic Large Magellanic Cloud (LMC) analog system, and was used in \cite{2020ApJ...893...48N} to fit the observed MW satellite galaxy population.} The images were made assuming the MW host halo center at an angular diameter distance of 1.75 $\mathrm{Mpc}\ h^{-1}$ from the observer, and the vertical aperture of the images is $15^{\circ}$. The images were created using the state-of-the-art multi-transparent-layer phase-space tessellation method described in~\citet{KAEHLER201768,KAEHLER2018}.

To demonstrate the consistency in the evolution history of individual (sub)halos, we select five different subhalos in the zoom-in simulation along with the MW host halo and match them to corresponding low-resolution counterparts in the parent boxes. Particularly, our selection of (sub)halos in the zoom-in consists of the MW host halo with $M_{\mathrm{Peak}}=10^{12.04}\,\mathrm{M_{\astrosun}}$, while our subhalos (numbered 1 to 5) have peak halo masses $\log_{10} (M_{\mathrm{Peak, z}}/\,\mathrm{M_{\astrosun}}) = \left[11.32, 10.90, 10.17, 9.12, 10.16\right]$, respectively. Our choice of the subhalos cover a variety of different dwarf galaxy regimes: Subhalo\ 1 is an LMC analog~\citep{2020ApJ...893...48N}, Subhalo 2 is another bright dwarf, Subhalo\ 3 is a bound (within $R_{\mathrm{vir}}$ of the MW host) classical dwarf dimmer than the LMC, Subhalo\ 4 is an ultra-faint dwarf (UFD) with no counterparts in the parent boxes, and Subhalo\ 5 is a splashback classical dwarf (i.e., a system previously inside but currently outside of the MW $R_{\mathrm{vir}}$, which appears to be inside $R_{\mathrm{vir}}$ in projection). Since we selected these subhalos based on their $M_{\mathrm{Peak}}$ values and classified them using the halo mass range in the left panel of Fig.~\ref{fig:1}, their stellar masses as labeled in the right column of Fig.~\ref{fig:3} do not strictly follow the definition in \citet{2017ARA&A..55..343B} due to scatter in the SMHM relation. Seen in the dark matter density maps in Fig.~\ref{fig:2}, the locations of the matched subhalos in the projected dark matter density maps agree well in the two parent boxes, but are noticeably different in the zoom-in, which could be explained by the differences in the orbital phases of subhalos across different resolutions as discussed below.

In Fig.~\ref{fig:3}, we present the evolution history of the six individual (sub)halos selected in the zoom-in along with their matched counterparts in the two parent boxes. The left column shows the evolution of $v_{\mathrm{max}}$ (a halo mass proxy free from pseudo-evolution due to cosmic expansion; e.g., \citealt{2006ApJ...647..201C,2013ApJ...766...25D}), the middle column shows the evolution of $R_{\mathrm{vir}}$ of the MW host and the 3D distance of the subhalos to the host $r_{\mathrm{3D}}$ in units of the MW virial radius $R_{\mathrm{vir, host}}$, and the right column shows the evolution of the intrinsic SFR values (uncorrected for observational effects) given by \textsc{UniverseMachine}. The triangles in the left and right panels with the same color as the evolution curves indicate the infall times (the first crossing of the MW host $R_{\mathrm{vir}}$, shown by the horizontal line in the middle column) of the subhalos in each resolution. The horizontal lines in the left column indicate the resolution limits in $v_{\mathrm{Mpeak}}$ that we estimate for the zoom-in ($10\ \mathrm{km\ s}^{-1}$), \textsc{c125-2048} ($35\ \mathrm{km\ s}^{-1}$), and \textsc{c125-1024} ($60\ \mathrm{km\ s}^{-1}$) boxes based on the analysis in Appendix~\ref{sec:A2}. 

For the MW host and Subhalos\ No.\ 1, 2, 3, and 5, the matched counterparts in the parent boxes have generally consistent $v_{\mathrm{max}}$ evolution features on timescales $\gtrsim 1$ Gyr. The history of Subhalo\ 4 (UFD) is well resolved in the zoom-in, but not so in the two parent boxes. Peaks in the $v_{\mathrm{max}}$ evolution curve generally indicate mergers, and this is captured well across the three resolution for the MW host, Subhalo\ 1, and Subhalo\ 2. The lower-mass Subhalo\ 3 (which is unresolved in \textsc{c125-1024}) and 5 (which is marginally resolved in \textsc{c125-1024}) show less correspondence in the short timescale variability of $v_{\mathrm{max}}$ across different resolutions. Since the intrinsic SFR at fixed $v_{\mathrm{max}}$ and redshift is mostly determined by $\Delta v_{\mathrm{max}}$, the SFR history in the right panel also shows short timescale differences (uncorrelated between time steps to model local variations in the ISM) in the three resolutions, while the long-term evolution agrees to first order as a result of the convergence in halo growth (long-term correlation dominated by the $v_{\mathrm{max}}$ evolution). 

We note that the growth of the MW host as well as the infall times of the subhalos in the zoom-in are systematically earlier than their counterparts in the parent boxes. The $R_{\mathrm{vir}}$ of the MW host in the zoom-in appears to suddenly increase due to the infall of Subhalo\ 1, which crosses the MW $R_{\mathrm{vir}}$ at $z\approx 0.06$. Although this could potentially bias the $r_{\mathrm{3D}}/R_{\mathrm{vir, host}}$ values of the zoom-in subhalos from row two to six, we have checked and confirmed that $r_{\mathrm{3D}}$ itself also evolves systematically earlier in the zoom-in. Given that the evolution of objects in \textsc{c125-2048} is not coherently ahead of their counterparts in \textsc{c125-1024}, the phenomenon of earlier-evolving zoom-in (sub)halos is not a pure resolution effect. Furthermore, it results in changes to the orbital phases and subhalo identity between the zoom-in and cosmological boxes (i.e., Subhalo\ 1 is within the MW virial radius in the zoom-in but has not accreted in the parent boxes, and Subhalo\ 5 is a splashback subhalo in the zoom-in but is found within the MW host halo's virial radius in the parent boxes). Since the UM DR1 model does not explicitly model environmental quenching using tidal or ram-pressure stripping, our results are only minutely affected by this artifact thanks to the agreement in the $v_{\mathrm{max}}$ evolution across the three resolutions. Nevertheless, future improvements of the UM DR1 model and other semi-analytical models of satellite galaxy evolution based on zoom-in simulations would be affected by this evolution systematic. We leave a careful numerical investigation of this issue to future work.

In summary, \textsc{UniverseMachine} applied to the \cite{2015ApJ...810...21M} MW zoom-ins and parent cosmological simulations agree well in different resolutions on a halo-to-halo basis for the halo ($v_{\mathrm{max}}$) and galaxy ($\mathrm{SFR_{raw}}$) growth, with the caveat that host halo $R_{\mathrm{vir}}$ growth and subhalo accretion times are systematically earlier in the zoom-ins. This difference does not significantly affect the main results in this work.

\section{Results}
\label{sec:4}

We now focus on the galaxy properties derived from applying \textsc{UniverseMachine} to zoom-ins, including the predicted SMHM relation and evolution (Section~\ref{sec:4.1}), star formation histories (Section~\ref{sec:4.2}), and quenched fraction of galaxies as a function of stellar mass (Section~\ref{sec:4.3}). We summarize the main takeaways in Section \ref{sec:4.4}. 

\subsection{Stellar Mass--Halo Mass Relation}
\label{sec:4.1}

\begin{figure*}[!htbp]
    \centering
	\includegraphics[width=2\columnwidth]{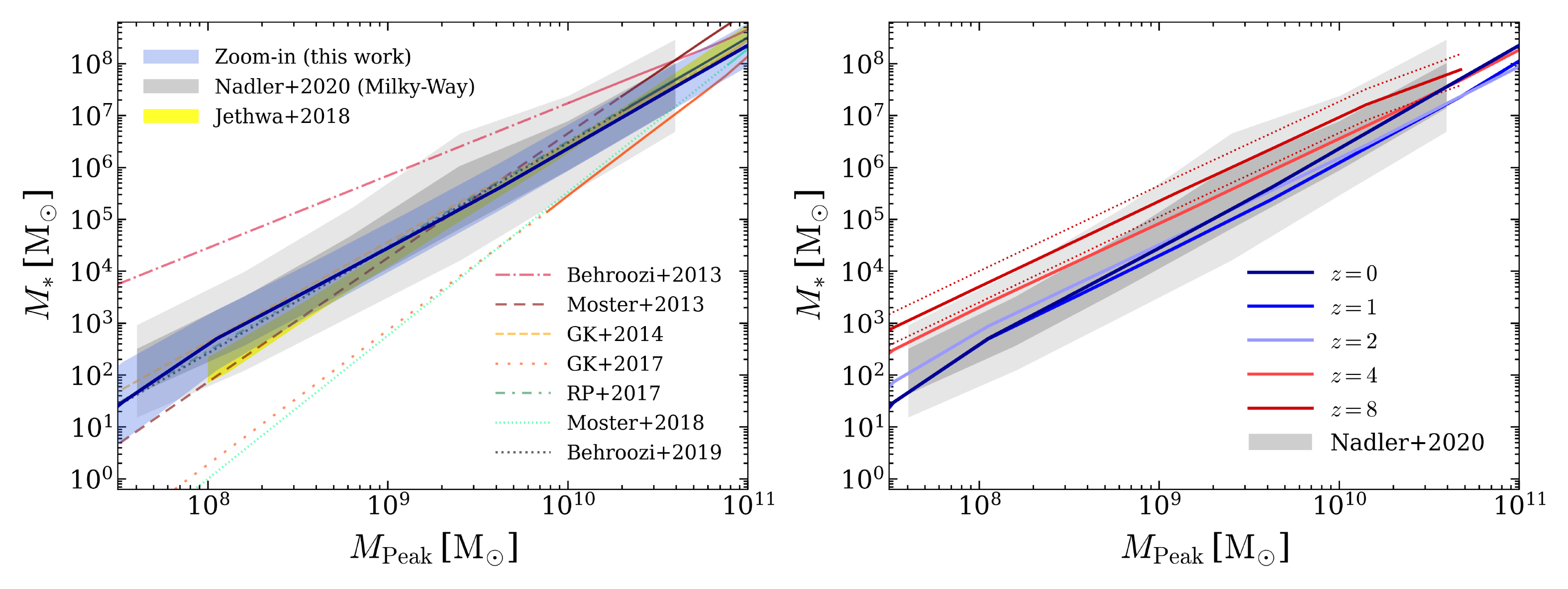}
    \caption{\emph{Left panel:} Low-mass end of the stellar mass--halo mass relation with \textsc{UniverseMachine} applied to the 45 MW zoom-ins (solid blue lines indicate the median relation and the shaded blue region indicates the 68\% scatter. Stellar mass--halo mass relations predicted by several recent models (at $z\sim0$) are shown for comparison (solid parts of the curves indicate the models' constrained mass ranges, see Section \ref{sec:4.1} for details). Our predictions are broadly consistent with these results and with the inferred stellar mass--halo mass relation for MW satellite galaxy observations using abundance matching (gray band; \citealt{2020ApJ...893...48N}). In addition, our prediction is consistent with a power-law extrapolation of UM DR1 predictions at higher masses, from B19. \emph{Right panel:} Evolution of the median stellar mass at fixed peak halo mass from $z=8$ to $z=0$. The median SMHM relations evolve by an order of magnitude from $z=8$ to $z=0$, which agree with the abundance matching-inferred SMHM relation of MW satellites within an uncertainty of $2\sigma$ throughout the evolution history. Dotted lines denote the $\pm 0.3$ dex (factor of 2) of the median SMHM relation at $z=8$. As discussed in Section~\ref{sec:5}, the true median $M_{\ast}$ at fixed $M_{\mathrm{Peak}}$ maybe larger by a factor $\sim2$ at $z=8$, leading to an even larger true median SMHM relation evolution, as UM tends to overestimate the SFRs of dwarf galaxies at $z\lesssim 4$.}
    \label{fig:4}
\end{figure*}

Fig.~\ref{fig:4} shows the low-mass end of the SMHM relation obtained by applying \textsc{UniverseMachine} to the 45 MW zoom-in simulations (solid blue line and shaded region), compared to other recent models and inferences of the low-mass SMHM relation. The SMHM relation predicted by running \textsc{UniverseMachine} on our zoom-in simulations is almost an exact power-law extrapolation of the high-mass-constrained UM DR1 model down to the zoom-in halo mass resolution limit ($\sim 5\times10^{7}\,\mathrm{M_{\astrosun}}$). 

The slope and scatter of the SMHM for the UM DR1 model applied to zoom-ins is in excellent agreement with  the SMHM relation inferred from state-of-the-art MW satellite observations using the Dark Energy Survey (DES) and Pan-STARRS combined with an abundance matching approach ~\citep{2020ApJ...893...48N}. This agreement might have been expected, given that \textsc{UniverseMachine} is tuned to match the global stellar mass function (down to $M_{\ast}\sim 10^{7}\,\mathrm{M_{\astrosun}}$), and \cite{2020ApJ...893...48N} find that the MW satellite luminosity function is consistent with a power-law extrapolation of the faint end of the GAMA luminosity function \citep{Loveday150501003}. However, this agreement does \emph{not} imply that the UM DR1 model captures all of the relevant physical processes that lead to this SMHM relation in the real Universe. As we discuss in Section~\ref{sec:5}, the predicted star formation histories (Section~\ref{sec:4.2}) and quenched fractions (Section~\ref{sec:4.3}) for dwarf galaxies indicate that the \textsc{UniverseMachine} faint-end SMHM relation evolves too strongly at late times.

We also compare our result for the SMHM relation to the results of several other models, including other abundance matching models~(\citealt{2014MNRAS.438.2578G,2017MNRAS.464.3108G}, labeled as GK14 and GK17)  informed by the Local Group satellite and field galaxy populations, as well as several recent empirical modeling results~(\citealt{2013ApJ...770...57B,2013MNRAS.428.3121M,2018MNRAS.477.1822M,2018MNRAS.473.2060J}, and \citealt{2017MNRAS.470..651R} labeled RP). The \textsc{UniverseMachine} SMHM prediction has a shallower slope at the low-mass end compared to models that introduce a large scatter in the faint-end SMHM relation~\citep{2017MNRAS.464.3108G}. This is encouraging, as recent observational analyses~\citep{2020ApJ...902..124C} find that the steep faint-end SMHM relation of ~\citet{2017MNRAS.464.3108G} tends to under-predict the abundance of bright satellites ($M_{\mathrm{Peak}}\gtrsim 10^{11}\,\mathrm{M_{\astrosun}}$) in the local universe. The UM DR1 model 
agrees reasonably well with most of models that are constrained at higher masses and extrapolated down to the dwarf galaxy mass range. Our result has a steeper SMHM relation slope than the \citet{2013ApJ...770...57B} slope extrapolated from higher masses, but is shallower than the ~\citet{2013MNRAS.428.3121M,2018MNRAS.477.1822M} slopes, which are close to \citet{2017MNRAS.464.3108G}.

As demonstrated in Section~\ref{sec:3.3} (Fig.~\ref{fig:3}), the dark matter halo assembly history is well resolved for objects in the MW zoom-in simulations up to redshift $z\sim 8$.\footnote{In particular, the $v_{\mathrm{max}}$ values of these halos are always above the $v_{\mathrm{Mpeak}}$ resolution floor as defined in Appendix \ref{sec:A2}.} Therefore, we can robustly trace the the SMHM relation back in time to visualize its evolution (Fig.~\ref{fig:4}, right panel). The median SMHM relation ($M_{\ast}$ at fixed $M_{\mathrm{Peak}}$) evolves over an order of magnitude from $z=8$ to $z=0$ for peak halo masses below $\sim 10^{9}\,\mathrm{M_{\astrosun}}$, and there is a mild trend for the slope to steepen with decreasing redshift. We note that the upper limit on the scatter in galaxy luminosity at fixed halo properties inferred from MW satellite observations \citep{2020ApJ...893...48N} can be used to inform the maximum degree of evolution allowed in the SMHM relation. In Sections~\ref{sec:4.4} and \ref{sec:5}, we discuss how the SMHM relation evolution is meaningful even though there are limitations in the star formation histories predicted by the UM DR1 model.

At all redshifts we consider, the SMHM relations for the zoom-ins shown here are consistent with direct power-law extrapolations of the high-mass end SMHM relations that result from applying \textsc{UniverseMachine} to the {\it Bolshoi-Planck} simulation (B19). This results from the fact that \textsc{UniverseMachine} scales the SFR of star forming galaxies as a double-power-law with $v_{\mathrm{max}}$, which reduces to a single power-law with respect to $M_{\mathrm{Peak}}$ at the low-mass end ($M_{\ast}\lesssim 10^{9}\mathrm{M_{\astrosun}}$, $M_{\mathrm{Peak}}\lesssim 10^{11}\mathrm{M_{\astrosun}}$). Naturally, the time integration of the SFR (evaluating $M_{\ast}(t)$) will preserve a power-law relation of the total stellar mass $M_{\ast}$ and the peak halo mass $M_{\mathrm{Peak}}$, while the change in the SMHM relation slope reflects the time evolution of the underlying halo growth rate. At fixed $v_{\mathrm{max}}$ (halo mass) and redshift, the scatter in SFR is determined by the instantaneous accretion rate of the halo ($\Delta v_{\mathrm{max}}$), which in turn leads to scatter in the SMHM relation at fixed halo mass. 

These results extend the peak halo mass range of the SMHM evolution from $10^{9}\,\mathrm{M_{\astrosun}}$ in \citet{2020MNRAS.499.5702B} to $10^{7.5}\,\mathrm{M_{\astrosun}}$. As shown in Figure 12 of \citet{2020MNRAS.499.5702B}, the trend of decreasing $M_{\ast}/M_{\mathrm{Peak}}$ at fixed $M_{\mathrm{Peak}}$ that we observe from $z=8$ to $z=2$ in the right panel of Fig.~\ref{fig:4} begins as early as $z\sim14$ and is mainly driven by the different evolution rates of halo and galaxy number densities at high redshifts. Upcoming JWST observations will significantly reduce the uncertainties in halo and galaxy number density evolution as early as $z\sim 10$, providing unprecedented constraints on the amplitude and evolution of the high-redshift SMHM relation. However, the progenitors of systems with present-day absolute UV magnitudes fainter than $M_{\mathrm{UV}}\sim -18$ are not observable at early times, even with JWST \citep{Boylan-Kolchin150406621}. Our framework for predicting galaxy growth at the faint end is therefore crucial for understanding the evolution of these extreme systems and its interplay with the onset, duration, and strength of reionization.

\subsection{Star Formation Histories}
\label{sec:4.2}

\begin{figure*}
    \centering
	\includegraphics[width=1.8\columnwidth]{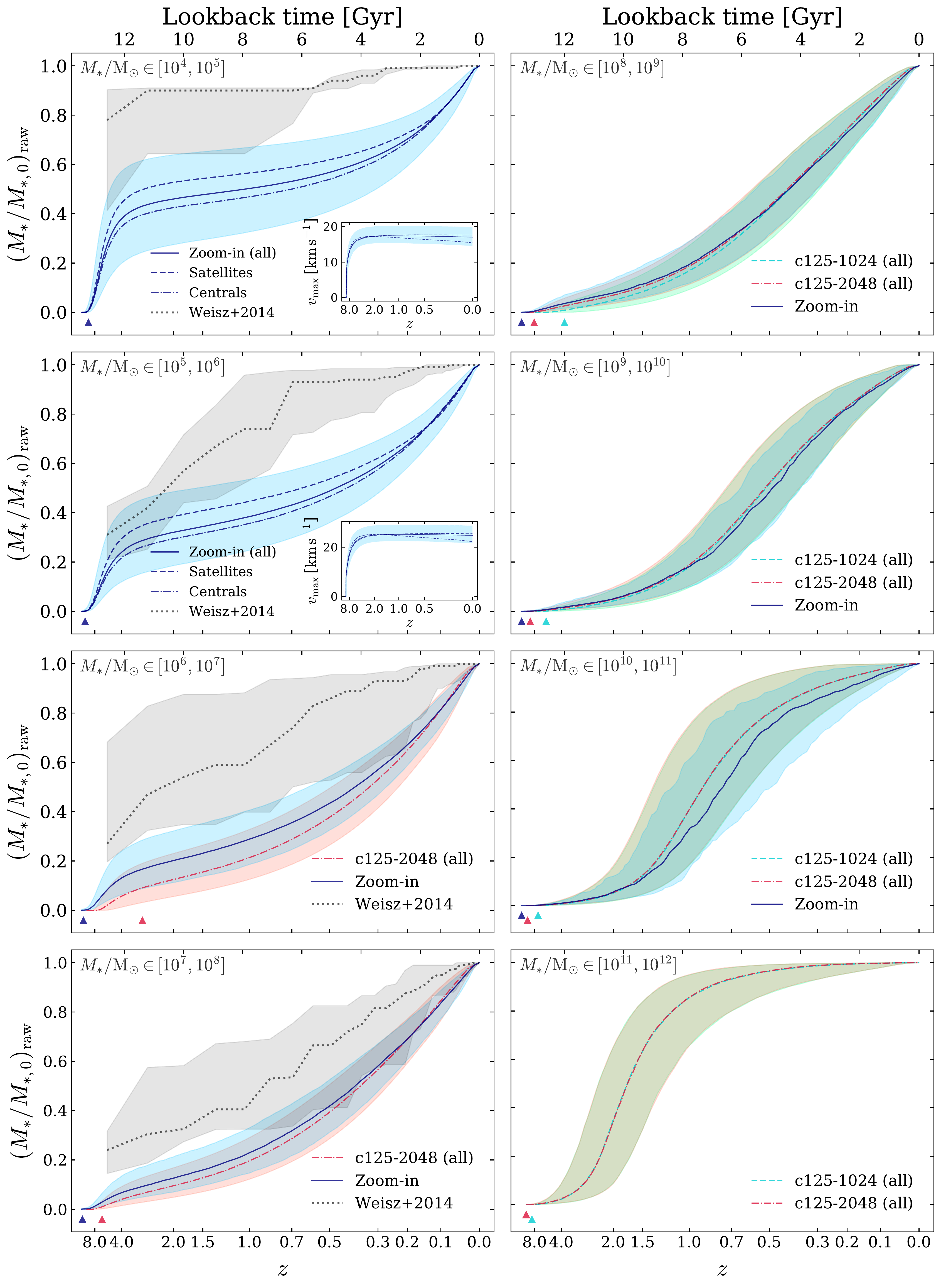}
    \caption{Cumulative star formation histories of galaxies in eight $z=0$ stellar mass bins derived by applying \textsc{UniverseMachine} to our three simulation resolution levels. Curves and shaded regions indicate the median and 68\% scatter of the SFH distribution for systems in our zoom-in simulations (blue) and the \textsc{c125-2048} (red) and \textsc{c125-1024} (green) parent boxes. In each stellar mass bin, we only include the simulation(s) that are well resolved and that sample the corresponding halo mass range. The observed median SFHs of dwarf galaxies in the Local Group~\citep{2014ApJ...789..147W} are shown for comparison in the left column (shaded grey regions indicate the 68\% scatter). We also show the median SFH for only central or satellite galaxies in the top two panels of the left column, as well as these halos' (non-orphans, $\gtrsim 93\%$ subhalos in these two mass bins) $v_{\mathrm{max}}$ evolution in the inset panels. Triangles of the same color as the solid curves indicate the redshift at which the median $v_{\mathrm{max}}$ of the subhalos for galaxies in a given resolution and stellar mass bin rises above the resolution limits (Appendix~\ref{sec:A2}).}
    \label{fig:5}
\end{figure*}

We break down the evolution of the SMHM relation discussed above by studying the star formation histories (SFH) for galaxies of different masses as predicted by \textsc{UniverseMachine}. The cumulative star formation histories (all stars ever formed in the galaxy, \emph{not} accounting for stellar mass loss, quantified by $M_{\ast}/M_{\ast, 0}$, the evolution of the galaxy stellar mass in units of their final $z=0$ stellar mass) for galaxies in eight different stellar mass bins are shown in Fig.~\ref{fig:5}. The galaxies binned based on their $z=0$ stellar mass, and we trace their total stellar mass evolution back along their \textsc{Consistent-Trees} accounting for their full merger history. In the inset panels, we only include galaxies from the simulations with well-resolved low-mass halos or sufficient statistics for high-mass halos. The median and 68\% scatter of the SFHs (modeled as all stars ever formed) inferred for dwarf galaxies in the Local Group~\citep{2014ApJ...789..147W} are shown in the left column panels for comparison. 

The SFH at the three resolution levels agree well in the high-stellar-mass bins for both the median and the scatter (right column). In the $M_{\ast}/\,\mathrm{M_{\astrosun}}\in [10^{10}, 10^{11}]$ bin, the zoom-in galaxies tend to systematically form stars later than their counterparts in the parent boxes. A factor that leads to this difference is that there are \emph{no} massive MW-mass satellites in the joint 45 MW zoom-ins (prevalent in hosts much more massive than the MW) compared to the parent boxes. We have checked that the consistency between the three resolutions improves in this stellar mass bin if we only include central galaxies from the two parent boxes. Another contribution to this difference might come from our selection of isolated MW-mass halos only, as a recent study using the FIRE simulation found earlier SFH for central galaxies in Local Group-like environments compared to isolated MW halos, due to earlier halo mass accretion in the former~\citep{2019MNRAS.489.4574G}.

For lower stellar mass galaxies (left column), we compare the joint zoom-in SFHs and the SFHs from the lower resolution \textsc{c125-2048} box. We also plot the observed SFHs of dwarf galaxies from \citet{2014ApJ...789..147W} (starting from $z\sim 5.5$, lookback time $\sim 12.59$ Gyrs). The zoom-in galaxies tend to show earlier star formation than their parent box counterparts when we go from the $M_{\ast}/\,\mathrm{M_{\astrosun}}\in [10^{7}, 10^{8}]$ bin to the $M_{\ast}/\,\mathrm{M_{\astrosun}}\in [10^{6}, 10^{7}]$ bin. Here, resolution effects kick in such that the zoom-in simulations resolve halo growth earlier than the parent box, while their late-time star formation converges at low redshift. Compared to the inferred SFHs of Local Group dwarf galaxies from ~\citet{2014ApJ...789..147W}, the rapid early star formation of the zoom-in dwarf galaxies at $z\gtrsim 3$ is marginally consistent with these observed galaxies. 

The UM DR1 model predicts that the zoom-in dwarf galaxies only form $\sim50\%$ of their present-day stellar mass in this early phase, and their star formation gets stalled by the slackened halo growth from $z\sim2$ to 1, before forming the other half of their stellar mass at low redshift ($z\lesssim 1$). This finding is valid for centrals and satellites alike, and is discrepant with the inferred late-time SFHs of dwarf galaxies in the Local Group within the similar stellar mass ranges, especially for ultra-faint dwarfs ($M_{\ast}\leqslant 10^{5}\,\mathrm{M_{\astrosun}}$) which are thought to quench their star formation at high redshifts ($z\gtrsim 5$) via reionization~(e.g.,~\citealt{2000ApJ...539..517B,2001PhR...349..125B,2002MNRAS.333..156B,2019MNRAS.483.4031R}). The differences of \textsc{UniverseMachine} with the Local Group observations in the ultra-faint dwarf regime is robust; their object-to-object scatter does not overlap, and it cannot be accounted for by observational uncertainties ($\lesssim 0.2$ in $M_{\ast}/M_{\ast, 0}$) at $z\lesssim 0.5$~\citep{2014ApJ...789..147W}. Even though updated stellar evolution models~(e.g.,~\citealt{2017ApJ...840...99D}) that account for atomic diffusion and gravitational settling could lead to slightly younger stellar ages, the tension between our SFHs and the Local Group dwarf SFHs cannot be fully resolved. It is not surprising that \textsc{UniverseMachine} predicts a different SFH than observed dwarfs since {\it no} constraints were placed on the cosmic star formation histories of low-mass galaxies ($M_{\ast} \lesssim 10^{8}\,\mathrm{M_{\astrosun}}$) in the fitting of the UM DR1 model. Similar late-time star formation that is discrepant from the Local Group dwarfs was predicted in \citet{2012ApJ...745..149L} when the observed SFR-$M_{\ast}$ relation was extrapolated down to the dwarf galaxy regime. 

In the top two panels of the left column in Fig.~\ref{fig:5}, we also show the evolution of $v_{\mathrm{max}}$ for all galaxies, centrals-only, and satellites-only samples in the two lowest SFH stellar mass bins. At $z\geqslant 4$, $v_{\mathrm{max}}$ increases steadfastly in both stellar mass bins and fuels the rapid early star formation due to halo growth. However, at $z\leqslant2$, $v_{\mathrm{max}}$ remains almost constant for centrals, and it decreases mildly on average for the subhalos hosting satellite galaxies due to tidal stripping after infall, which leads to an almost non-evolving $v_{\mathrm{max}}$ distribution for all galaxies in both stellar mass bins. This suggests that the late-time star formation observed at $z\lesssim 0.5$ in these low-mass bins is {\it not} driven by (sub)halo dark matter accretion at late times. Instead, the late-time star formation in these small halos results from a significant increase in the efficiency of their star formation while little change occurs in their halo mass. 

This phenomenon is due to the specific form of coupling that the \textsc{UniverseMachine} model assumes between SFR and $v_{\mathrm{max}}$, which behaves unreasonably when extrapolated to the low halo masses we examine here. The \textsc{UniverseMachine} model parameterizes the SFR coupling to $v_{\mathrm{max}}$ of star-forming galaxies in the form of a broken power-law (see Equation 4 in B19):
\begin{equation}
    \label{eq:3}
    \mathrm{SFR_{SF}} \propto \left[ \left(\frac{v_{\mathrm{max}}}{v_{0}}\right)^{\alpha} + \left(\frac{v_{\mathrm{max}}}{v_{0}}\right)^{\beta}\right]^{-1}
\end{equation}
where $v_{0}$, $\alpha$, $\beta$ are model parameters, and we have assumed $v_{\mathrm{Mpeak}}\approx v_{\mathrm{max}}$ since $v_{\mathrm{max}}$ is almost constant at $z\leqslant 2$. The power-law slopes $\alpha$ and $\beta$ account for the different star formation efficiencies in low- and high-mass galaxies due to different feedback and quenching mechanisms, whereas the normalization $v_{0}$ sets the location of that transition. The evolution of the $\alpha$ and $\beta$ values in the UM DR1 model are mostly constrained by the decreasing specific SFR (star formation efficiency) of observed galaxies with $M_{\ast}\geqslant 10^{9}\,\mathrm{M_{\astrosun}}$ (see Figure 3 right panel in B19), which corresponds to halo masses of $M_{\mathrm{Peak}}\geqslant 2\times 10^{11}\,\mathrm{M_{\astrosun}}$. The underlying physics here encompasses the AGN feedback~\citep{2003ApJ...595..614W,2005Natur.433..604D,2005MNRAS.361..776S,2012ARA&A..50..455F,2013ARA&A..51..511K} and merger-driven quenching in massive galaxies~\citep{2009ApJ...703.1531N,2009ApJ...706L..86N,2013ApJ...766...71R,2016MNRAS.456.1030W}, which becomes more effective in massive halos at low redshift and reduces the efficiency at which a halo can convert its gas into stars. 

The best-fit UM DR1 model values for ($v_{0}$, $\alpha$, $\beta$) are ($137.72\,\mathrm{km\,s^{-1}}$, $-6.135$, $-1.813$) at $z=0$. At $z\lesssim 0.5$, $v_{0}$ remains almost constant, while both $\alpha$ and $\beta$ increase steadily towards $z=0$. For the galaxies in the three highest stellar mass bins ($M_{\ast}\geqslant 10^{9}\,\mathrm{M_{\astrosun}}$) in Fig.~\ref{fig:5}, most of their host (sub)halos satisfy $v_{\mathrm{max}}\gtrsim 120\,\mathrm{km\,s^{-1}}$, i.e. $(v_{\mathrm{max}}/v_{0})\gtrsim 1$. With the parameterization in Equation~\ref{eq:3}, even if $v_{\mathrm{max}}$ remains almost constant at $z\lesssim 0.5$ for these massive galaxies, the UM DR1 model predicts that SFRs {\it decrease} with time towards $z=0$ due to the increase in $\alpha$ and $\beta$ values, which is evident the right column of Fig.~\ref{fig:5}.

\begin{figure}[h!]
    \centering
	\includegraphics[width=\columnwidth]{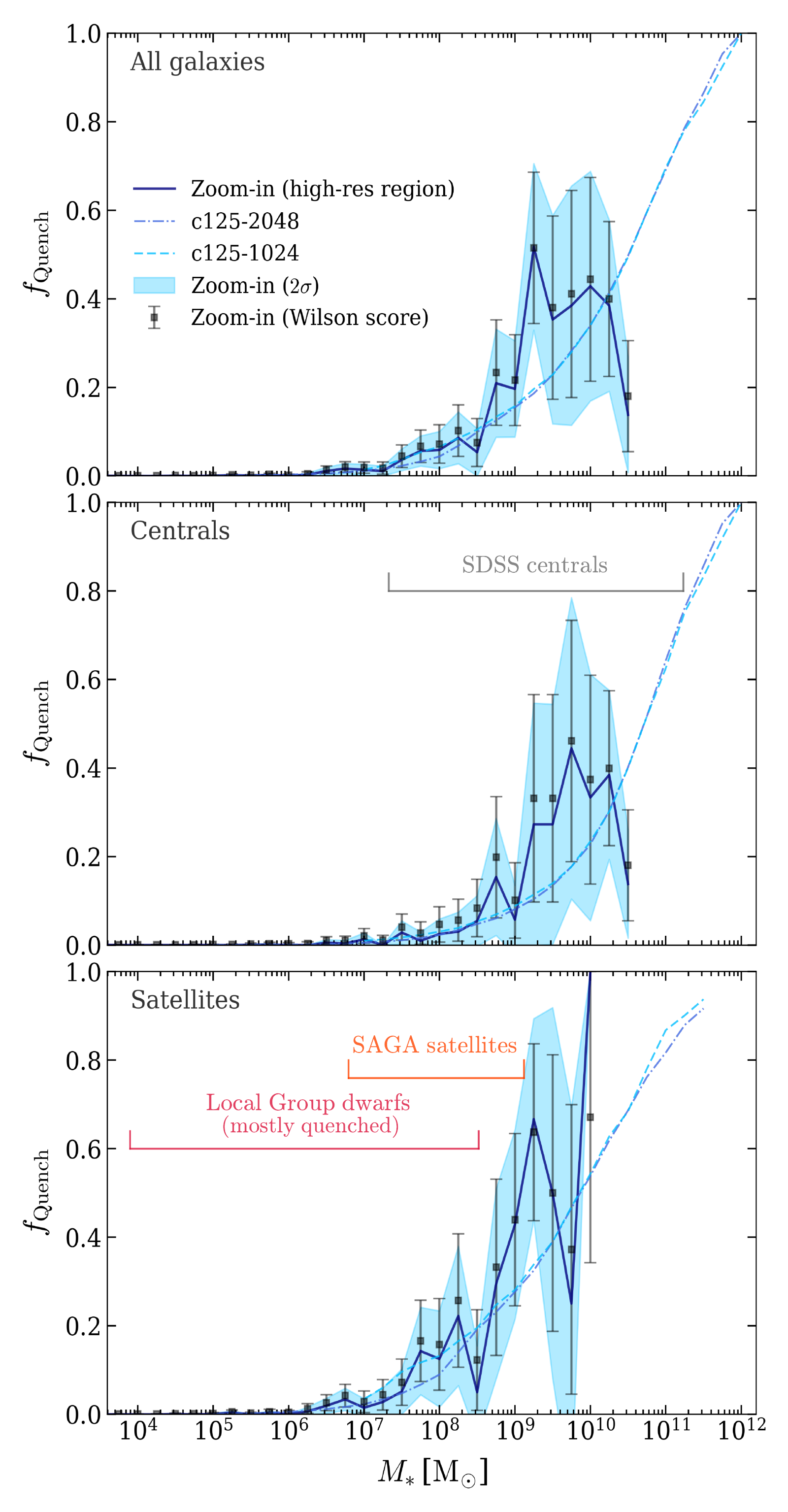}
    \caption{ Quenched fraction as a function of stellar mass for all galaxies (top), only centrals (center), and only satellites (bottom) in the joint MW resimulations (solid blue), \textsc{c125-2048} (dot-dashed blue), and \textsc{c125-1024} (dashed blue). The zoom-in values were derived using only galaxies within $r_{90}$ (Section~\ref{sec:3.1}) of each MW host. The shaded regions indicate the $2\times$ standard deviation of 500 bootstrap samples, while the black error bars represent the $95\%$ confidence Wilson score interval in each stellar mass bin for the zoom-in galaxies. The {\it mass ranges} of observational datasets (see text for details) that could be used for future model improvements are indicated in the lower two panels. }
    \label{fig:6}
\end{figure}

However, directly extrapolating this model down to lower-mass galaxies ($M_{\ast} < 10^{9}\,\mathrm{M_{\astrosun}}$) could result in $(v_{\mathrm{max}}/v_{0}) < 1$, which would lead to {\it increasing} SFR values from $z\sim 0.5$ to $z=0$ due to increasing values of $\alpha$ and $\beta$ even though $v_{\mathrm{max}}$ remains constant at low redshift. This effect can be seen in Fig.~\ref{fig:5} (left panel) and becomes stronger as stellar mass decreases. As the model predicts all galaxies to be star forming below $M_{\ast}\sim 10^{7}\,\mathrm{M_{\astrosun}}$ (see Section~\ref{sec:4.3} and Fig.~\ref{fig:6}), the late-time star formation is evident for both centrals and satellites. This effect is smaller for satellites due to the mild decrease of their $v_{\mathrm{max}}$ at $z\leqslant 2$, due to tidal stripping after being accreted. 

These findings reflect the fact that the global stellar mass function extends smoothly from $M_{\ast}\gtrsim 10^{9}\,\mathrm{M_{\astrosun}}$ down to $M_{\ast}\lesssim 10^{5}\,\mathrm{M_{\astrosun}}$, leading to a realistic SMHM relation (Fig.~\ref{fig:4}). However, extending the SFR($v_{\mathrm{max}}$) coupling constrained by high-mass-galaxy star formation efficiency, where quenching is driven by AGN feedback and mergers, does not fit in the dwarf galaxy regime, where quenching is driven by stripping, stellar feedback, and reionization \citep[e.g.,][]{2000ApJ...539..517B,2001PhR...349..125B,2002MNRAS.333..156B,2019MNRAS.483.4031R}), and results in the unrealistic SFHs at $z\lesssim 1$ (Fig.~\ref{fig:5}, left column). We discuss the implications of model improvements based on this discrepancy along with the quenched fractions (next section) predicted by \textsc{UniverseMachine} for dwarf galaxies in Section~\ref{sec:5}.

\subsection{Quenched Fractions}
\label{sec:4.3}

Another crucial observational constraint used to constrain the UM DR1 model is the quenched fraction ($f_{\mathrm{Quench}}$) of galaxies as a function of stellar mass, which sets the transition from the star-forming main sequence to the quenched red sequence of galaxies~\citep{2006MNRAS.368....2D,2007ApJ...665..265F,2007ApJ...660L..43N}. The definition of ``quenched'' in the model is given by a set of redshift-dependent criteria that depend on both the specific star formation rate (SFR per unit stellar mass, hereafter sSFR; \citealt{2013ApJ...767...50M}) and UVJ-band colors~\citep{2013ApJ...777...18M}. We follow the definition in Appendix C6 of B19 and adopt $\mathrm{sSFR < 10^{-11} yr^{-1}}$ as the criterion for classifying quenched galaxies at $z=0$.

The $f_{\mathrm{Quench}}$-$M_{\ast}$ relations for all galaxies, centrals, and satellites are shown from in Fig.~\ref{fig:6}. The classification of galaxies into `star forming' or `quenched' forms a binomial distribution in each stellar mass bin. We show $2\sigma$ bootstrap errors (from 500 re-samplings), as well as the $95\%$ confidence Wilson score interval~\citep{Wilson1927} for the three categories of zoom-in galaxies. For brevity, the small bootstrap standard deviations in the two parent boxes are not shown due to the abundant sampling of galaxies at the well-resolved halo masses. The three resolutions agree in the overall trend of the $f_{\mathrm{Quench}}$-$M_{\ast}$ relation, although the zoom-in results show significant scatter towards high masses due to limited statistics. Similar to the $M_{\ast}/\,\mathrm{M_{\astrosun}}\in [10^{10}, 10^{11}]$ bin in Fig.~\ref{fig:5}, the zoom-ins begin to miss out on MW-mass satellites and lead to an $f_{\mathrm{Quench}}$ value that is biased high, although the Wilson score interval remains consistent at the $2\sigma$ level with the parent boxes. This also leads to the spike at $\sim 10^{10}\,\mathrm{M_{\astrosun}}$ in the top panel when combined with the centrals.

The $f_{\mathrm{Quench}}$-$M_{\ast}$ relation including both central and satellite galaxies in the top panel of Fig.~\ref{fig:6} suggests that the $f_{\mathrm{Quench}}$-$M_{\ast}$ relation is a simple extension of the UM DR1 model constrained at higher masses down to the ultra-faint dwarf regime in the form of an error function (see Section 3.2 in B19). Comparing the lower two panels, the satellites generally have higher quenched fractions at fixed stellar mass compared to centrals above $M^{\ast}\sim10^{7}\,\mathrm{M_{\astrosun}}$ in both the zoom-ins and the parent boxes. However, the classical and ultra-faint dwarfs below this mass are all predicted to be star forming regardless of their central or satellite identity. Recent observations~\citep{2012AJ....144....4M,2013AJ....146...46K,2015ApJ...804..136W,2021ApJ...907...85M} suggest that satellite galaxies around MW-mass hosts in the Local Group and low-redshift Universe are at least $\sim 60\%$ quenched at $\sim 10^{7}\,\mathrm{M_{\astrosun}}$, with higher quenched fractions towards lower stellar masses. This trend around MW-hosts is also reproduced in hydrodynamic zoom-in simulations~\citep{2018MNRAS.478..548S,2020arXiv200802805A}, although these simulations have their own issues of over-quenching field galaxies compared to observations~\citep{2020arXiv201001132D}.  We indicate the mass ranges of specific datasets (field centrals, \citealt{2012ApJ...757...85G}; SAGA Survey satellites, \citealt{2021ApJ...907...85M}, and Local group dwarfs, \citealt{2014ApJ...789..147W}) that could be used to constrain future \textsc{UniverseMachine} model improvements. 

The discrepancy between \textsc{UniverseMachine} and observations reveals that extending the error function-like parameterization of $f_{\mathrm{Quench}}$-$M_{\ast}$ for higher-mass galaxies to lower stellar masses is insufficient to capture the correct satellite quenched fractions in the dwarf regime. Because the UM DR1 model predicts that \emph{all} low-mass galaxies are star forming, a more realistic treatment of observational selection functions is not expected to ease this tension. This implies that a different parameterization for the dwarf galaxy quenched fractions is necessary, in addition to the single-error-function module that fully described the observational constraints at $M_{\ast}\gtrsim 10^{9}\,\mathrm{M_{\astrosun}}$ in B19. As a consequence of not quenching dwarf galaxies with $M_{\ast}\lesssim 10^{7}\,\mathrm{M_{\astrosun}}$, the DR1 model fails to decouple SFR from halo growth scaled as Equation~\ref{eq:3}, and forces these low-mass galaxies into a late star forming epoch ($z\lesssim 1$) to match the global stellar mass function at different redshifts, which would have been otherwise prevented if most of these dwarfs were quenched (quenched galaxies in \textsc{UniverseMachine} takes on a median SFR of $10^{-11.8}\,\mathrm{yr^{-1}}$ with a 0.3 dex scatter, regardless of halo mass or halo accretion rate). 

\subsection{Discussion}
\label{sec:4.4}

Summarizing the main results of this section, \textsc{UniverseMachine} predicts a low-mass SMHM relation that is a power-law extrapolation of the SMHM relation at $M_{\ast}\geqslant 10^{9}\,\mathrm{M_{\astrosun}}$. This relation is already in place at $z=8$ and evolves over an order of magnitude to a slope and scatter consistent with abundance matching-inferred MW satellite SMHM relation at $z=0$. The consistency in the SMHM relation with the SMHM relation inferred from MW satellite observations~\citep{2020ApJ...893...48N} follows from the fact that the UM DR1 model matches global stellar mass functions across different redshifts. However, as the SFR($v_{\mathrm{max}}$) coupling is mostly constrained by the specific SFR evolution of high-mass galaxies ($M_{\ast}\gtrsim 10^{9}\,\mathrm{M_{\astrosun}}$), the model forces all dwarf galaxies ($M_{\ast}\lesssim 10^{7}\,\mathrm{M_{\astrosun}}$) to be actively star forming below $z\sim1$, regardless of their central or satellite identity. As a consequence, the path through which the dwarfs evolve onto this SMHM relation is inconsistent with that for observed dwarf galaxies.

Although the dwarf galaxy star formation histories predicted by \textsc{UniverseMachine} are not realistic, the predicted evolution of the SMHM relation is reasonable given that \textsc{UniverseMachine} matches the global stellar mass function down to $M_{\ast}\sim 10^{7}\mathrm{M_{\astrosun}}$. As shown in the right panel of Fig.~\ref{fig:4}, the SMHM relation evolves by an order of magnitude in stellar mass at fixed $M_{\mathrm{Peak}}$ from $z=8$ to $z=0$. Even if we implement improvements to the DR1 model to enforce SFHs that resemble those for observed dwarf galaxies, stellar masses at $z>4$ are only required to change by factor of $\sim 2$ at these high redshifts due to the very rapid early-phase star formation (Fig.~\ref{fig:5} left column). Therefore, future improvements to the predicted dwarf galaxy SFHs will not significantly affect the predicted faint-end SMHM evolution.

Preliminary results using the predicted SFHs in Section~\ref{sec:4.2} indicate that the present-day UFDs' progenitors were \emph{not} the major contributors ($\sim 5\%$) to the ionizing-photon budget during reionization. This subdominant role is consistent with empirical modeling that connects Local Group dwarf galaxies to high-redshift progenitors based on their abundances and star formation histories \citep{2015MNRAS.453.1503B,2017MNRAS.469L..83W}. Since the current UM model may under-predict SFRs during reionization (more discussion in Section~\ref{sec:5}), this predicted reionization contribution may slightly increase in future work but is expected to remain subdominant.


Thus, \textsc{UniverseMachine} arrives at a faint-end SMHM relation in agreement with the SMHM relation inferred from MW satellites (\ref{sec:4.1}) despite its unrealistic predictions for dwarf galaxy star formation histories (\ref{sec:4.2}) and quenched fractions (\ref{sec:4.3}). This implies that the DR1 model does not fully capture the physical processes that govern dwarf galaxy evolution, which is unsurprising given that it was only constrained using data for higher-mass galaxies. We now discuss how \textsc{UniverseMachine} might be improved to account for the physics that quenches star formation in the low-mass regime.

\section{Implications for Model Improvements}
\label{sec:5}

The results in the previous section demonstrate specific shortcomings of the UM DR1 model when applied to lower mass objects that previously investigated. In particular, although \textsc{UniverseMachine} predicts an SMHM relation in good agreement with that inferred from abundance matching and other empirical models, it arrives at this SMHM relation in an unrealistic manner, with all dwarf galaxies actively forming stars at $z=0$. To more fully constrain the diverse range of physical processes that influence galaxy formation and evolution, \textsc{UniverseMachine} and other empirical models must self-consistently predict both the slope and scatter of the SMHM relation and the detailed evolution of particular galaxies along this relation over time.

Given the factors summarized in Section \ref{sec:4.4}, further improvements of the UM DR1 model call for at least implicit incorporation of environmental quenching effects to produce nonzero $f_{\mathrm{Quench}}$ values for dwarf satellites, thereby decoupling their SFR from halo growth at late times. This change would in principle mostly affect classical and bright dwarfs  ($M_{\ast}/\,\mathrm{M_{\astrosun}}\in [10^{5}, 10^{9}]$) in a mass-dependent fashion, and should also accommodate the short quenching timescale ($\lesssim2$ Gyrs) inferred for galaxies with $M_{\ast}\lesssim 10^{8}\mathrm{M_{\astrosun}}$ from Local Group dwarfs~\citep{2015ApJ...808L..27W}. Meanwhile, for ultra-faint dwarfs ($M_{\ast}/\,\mathrm{M_{\astrosun}} \leqslant 10^{5}$), an effective model that captures quenching due to reionization physics (e.g.,, UV heating which prevents gas cooling) could likewise decouple dark matter accretion and star formation. Implementing such a model and constraining it by comparing to the early-quenching SFHs of Local Group dwarf galaxies will offer insights into the timing and efficiency of reionization quenching. Both the environmental and the reionization quenching effects can potentially be modeled by an $f_{\mathrm{Quench}}$-$M_{\ast}$ parameterization that increases with decreasing stellar mass for $M_{\ast} \lesssim 10^{9}\,\mathrm{M_{\astrosun}}$, opposite to the parameterization at higher masses ($M_{\ast} \gtrsim 10^{9}\,\mathrm{M_{\astrosun}}$) in B19.

With these two changes implemented, the stellar masses of dwarf galaxies could be underestimated due to the removal of late-time ($z\lesssim 1$) star formation. To maintain a SMHM relation that matches constraints from Local Group dwarf galaxies, these changes necessitate that low-mass (sub)halos form stars more efficiently than high-mass halos at high redshifts ($z\gtrsim 5$), superseding the fiducial scaling of SFR with halo growth assumed in the UM DR1 model. This is consistent with the efficient high-redshift star formation inferred from Local Group dwarf galaxies~\citep{2014ApJ...790L..17M}. However, even if we consider the ultra-faint dwarfs that should be most affected by reionization and quench before $z\sim4$, they already form $\sim 50\%$ of their present-day stellar mass in the early star formation phase predicted by the UM DR1 model due to rapid halo growth. Thus a slight increase (a factor of few) in the early SFR would be sufficient to reproduce SFH consistent with \citet{2014ApJ...789..147W} while largely maintaining the SMHM relation. Additional information for the galaxy stellar age distribution as a function of halo mass may also provide quantitative constrains on the star formation efficiency of dwarf (sub)halos relative to their more massive counterparts at high redshift~\citep{2020arXiv201010520K}.

As discussed in Section~\ref{sec:4.4}, the current UM model predicts that local UFDs' progenitors contributed only $\sim5\%$ to the total ionization budget during reionization. This may represent a slight underestimate due to the finite resolution of our simulations and the current UM model under-predicting UFD star formation efficiencies at $z>4$. However, even with improved modeling and increased resolution, we anticipate that the UFD progenitors' contribution to reionization will be $\lesssim 10\%$ as the required increase in SFR to match Local Group SFHs is only factor $\sim 2$ at $z>4$. In order to make robust predictions for the UFD contribution to reionization, it is necessary to account for systematics in the mass dependence of the IMF as well as the spatial and temporal stochasticity of the UV photon escape fraction. These effects will be investigated further in future work.

In all, more accurate predictions of the earliest star formation epoch in the Universe given by UM will rely on incorporating dwarf galaxy constraints to the model. To emphasize, we advocate that SFH and $f_{\mathrm{Quench}}$ are essential factors that impact the galaxy--halo connection at the low-mass end, and empirical models that build upon the \textsc{UniverseMachine} framework need to carefully account for their mass-dependent behavior to predict galaxy evolution consistent with the real universe. Although we do not attempt to directly constrain the \textsc{UniverseMachine} model on dwarf galaxy scales in this work, we summarize two future directions for this work motivated by our current findings:
\begin{itemize}
    \item \emph{Re-fitting the \textsc{UniverseMachine} model}: We plan to carry out a detailed exploration of the UM DR1 model posterior space, including a study of parameter covariances as well as the flexibility of the current model to allow for early star formation in dwarf galaxies while maintaining appropriate scaling relations at higher masses (e.g.,, stellar mass functions, cosmic star formation history, quenched fractions, and auto- and cross-correlations). 
    
    \item \emph{Implementing dwarf galaxy quenching mechanisms}: In parallel, we plan to explore effective models of environmental quenching and reionization by implementing a more sophisticated model of the redshift-dependent $f_{\mathrm{Quench}}$-$M_{\ast}$ relation. Environmental quenching could potentially be constrained with observed quenched fractions in SDSS field centrals~\citep{2012ApJ...757...85G} and SAGA+Local Group satellites~\citep{2014ApJ...789..147W,2021ApJ...907...85M} hand in hand, while the impact of reionization could potentially be constrained by Local Group ultra-faint dwarfs~\citep{2014ApJ...789..147W}. This approach can be generalized to improve other aspects of the model at low masses that are inflexible due to the current parameterization based on high-mass galaxies only.
\end{itemize}

In addition to these plausible improvements, upcoming $z\sim 10$ JWST observations will significantly reduce uncertainties on the high-redshift SMHM relation for the progenitors of the brightest dwarf galaxies \citep{2020MNRAS.499.5702B}. Boosting the SFRs of dwarf galaxies at early times compared to the UM DR1 model as required by this study would suggest an increase the detectability of these systems at redshifts, thereby impacting predictions for JWST. Combined with our methodology to extend galaxy evolution predictions down to the ultra-faint dwarf galaxy regime, these high-redshift observations will further constrain the strength and timing of reionization, which holds far-reaching consequences for our understanding of dwarf galaxy formation and evolution.

\section{Conclusions and Outlook}
\label{sec:6}

We applied the empirical galaxy--halo connection model \textsc{UniverseMachine}~\citep{2019MNRAS.488.3143B} to cosmological zoom-in simulations of MW-mass halos along with their parent boxes. Our study is based on a joint sample of 45 MW halo zoom-in re-simulations in the halo mass range of $10^{12.1\pm0.03}\,\mathrm{M_{\astrosun}}$ selected from the parent cosmological boxes of the Chinchilla simulations (\textsc{c125-1024} and \textsc{c125-2048}, \citealt{2015ApJ...810...21M}). This work provides a framework for studying the galaxy--halo connection over the full range of observed galaxy masses, and it exposes specific aspects of the UM DR1 model that break down in the dwarf galaxy regime. The main findings of this work can be summarized as follows:
\begin{itemize}

    \item We have applied the empirical galaxy--halo connection model \textsc{UniverseMachine} to a set of 45 MW-mass zoom-in simulations, achieving seven decades of halo mass resolution and modeling star formation histories for the full range of galaxy masses when combined with their parent cosmological simulations (Fig.~\ref{fig:1}).
    
    \item The UM DR1 model predicts a SMHM relation that is consistent with observational constraints from MW satellites at $z=0$, including for ultra-faint dwarf galaxies (Fig.~\ref{fig:4}, left panel).
    
    \item Our predicted SMHM relation evolves roughly an order of magnitude from $z=8$ to $z=4$, and evolves more mildly from $z=4$ to the present (Fig.~\ref{fig:4}, right panel).
    
    \item The UM DR1 model predicts significant late-time star formation in dwarf galaxies (a quenched fraction near $0\%$ for $M_{\ast} \lesssim 10^{7} \,\mathrm{M_{\astrosun}}$), and is discrepant with the star formation histories and quenched fractions inferred for Local Group dwarfs (Fig.~\ref{fig:5}--\ref{fig:6}).
    
    \item We identify modifications to \textsc{UniverseMachine}, including environmental and reionization-driven quenching mechanisms for dwarf galaxies, that must be implemented in specific stellar mass regimes ($\lesssim 10^{7}\mathrm{M_{\astrosun}}$). Such models may simultaneously yield star formation history and SMHM relation predictions in agreement with dwarf galaxy observations (Section~\ref{sec:5}).

\end{itemize}

This work lays crucial methodological foundations (especially for ranking subhalos, Section~\ref{sec:3.3} and Appendix~\ref{sec:A1}) for applying \textsc{UniverseMachine} to zoom-in simulations and constraining it at the low-mass end. Our work demonstrates that scaling galaxy SFR with (sub)halo dark matter accretion rate can yield reasonably realistic stellar mass--halo mass relations and match the evolution of the global stellar mass function down to the ultra-faint dwarf galaxy regime. However, as discussed in Sections~\ref{sec:4} and \ref{sec:5}, our findings also reveal that this comes at the cost of predicting significant late-time star formation in dwarfs and not quenching the dwarf satellites at low redshifts ($z\lesssim 1$). These limitations arise due to the unconstrained nature of the UM DR1 model in the dwarf galaxy regime. Identifying these limitations in the current version of the \textsc{UniverseMachine} model demonstrates that it is necessary to model quenching differently in low-mass dwarfs ($M_{\ast} \lesssim 10^{7} \,\mathrm{M_{\astrosun}}$) than in their more massive counterparts ($M_{\ast} \gtrsim 10^{9} \,\mathrm{M_{\astrosun}}$), marking a significant transition of the underlying quenching physics with halo mass. As a major focus of upcoming work, a more comprehensive picture that captures the life cycle of star formation in dwarf galaxies will be built using detailed modeling of environmental and reionization quenching mechanisms as well as exploring the flexibility of the UM DR1 model to encompass diverse SFHs in dwarfs.

Apart from these two pertinent directions, comparison with hydrodynamic simulations (see ~\citealt{2020NatRP...2...42V} for a review) on both cosmological and zoom-in scales could also provide valuable information. For example, recent results~\citep{2019MNRAS.489.4574G} from the FIRE Simulation reproduce the stellar mass function and star formation histories of dwarf galaxies around MW-mass hosts reasonably well, while showing minimal environmental dependence in their SFHs. Comparing \textsc{UniverseMachine} to these results can shed light on the details regarding the diversity of SFHs and quenching processes given their power to self-consistently follow the baryonic evolution in galaxies with hydrodynamics.

Another prospect of our future work is to account for the systematic biases introduced into the stellar mass--halo mass relation by stochastically populating the stellar initial mass function (IMF) in ultra-faint dwarfs~\citep{2020MNRAS.492....8A}. Determining the origin of the earlier evolution of subhalos (Figs.~\ref{fig:2} and \ref{fig:3}) in zoom-in simulations is also crucial for understanding the systematics of environmental quenching effects due to the change in subhalo orbital phases.

The methodology we have developed can easily be generalized to other mass scales (e.g.,\ using cluster scale zoom-in re-simulations), enabling empirical studies of the galaxy--halo connection that benefit from the dynamic range achieved by zoom-in simulations. We have demonstrated that combining the \textsc{UniverseMachine} model with zoom-in dark matter simulations offers a powerful, high-resolution approach to study the relation between dark matter accretion and star formation. We expect that this methodology, combined with the model improvements highlighted here and with new observations that inform the abundance and evolution of dwarf galaxies, will play a key role in future studies of the galaxy--halo connection over large dynamic ranges.

\section*{Acknowledgements}

We thank the anonymous reviewer for valuable comments that improved the work. We thank Matt Becker for his work on the Chinchilla simulation suite used in this work. We thank Ralf Kaehler for making the brilliant rendition of the projected dark matter density maps shown in Fig.~\ref{fig:2}. We thank Pratik Gandhi, Andrey Kravtsov, Phil Mansfield, Dan Weisz, and Andrew Wetzel for helpful discussions and comments on the manuscript. YW acknowledges the support of a Stanford-KIPAC Chabolla Fellowship. This research received support from the National Science Foundation (NSF) under grant No.\ NSF DGE-1656518 through the NSF Graduate Research Fellowship received by EON, by the U.S. Department of Energy contract to SLAC No. DE-AC02-76SF00515, and by Stanford University. Y-YM\ is supported by NASA through the NASA Hubble Fellowship grant No.\ HST-HF2-51441.001 awarded by the Space Telescope Science Institute, which is operated by the Association of Universities for Research in Astronomy, Inc., under NASA contract NAS5-26555. SA's research was partially supported by the U.S. Department of Energy (DOE) Office of Science Distinguished Scientist Fellow Program. PB was partially funded by a Packard Fellowship, Grant \#2019-69646. This research made use of computational resources at SLAC National Accelerator Laboratory, a U.S.\ Department of Energy Office, and the Sherlock cluster at the Stanford Research Computing Center (SRCC); the authors are thankful for the support of the SLAC and SRCC computing teams.  This research made extensive use of \href{https://arXiv.org}{arXiv.org} and NASA's Astrophysics Data System for bibliographic information.

\bibliography{sfmw}{}
\bibliographystyle{aasjournal}

\appendix
\counterwithin{figure}{section}

\section{Subhalo ranking}
\label{sec:A1}

As briefly discussed in Section~\ref{sec:3.3}, correctly ranking subhalos in zoom-in simulations is non-trivial and is a crucial step for applying \textsc{UniverseMachine}. This involves mapping the halo accretion rate $\Delta v_{\mathrm{max}}$ to galaxy SFR at fixed halo mass ($v_{\mathrm{max}}$). The definition of $\Delta v_{\mathrm{max}}$ follows from Equation~1 in B19:
\begin{equation}
    \label{eq:A1}
    \Delta v_{\mathrm{max}} = \frac{v_{\mathrm{max}}(z_{\mathrm{now}})}{v_{\mathrm{max}}(\mathrm{max}(z_{\mathrm{dyn}}, z_{\mathrm{Mpeak}}))}\,,
\end{equation}
where $z_{\mathrm{now}}$ is the redshift when $\Delta v_{\mathrm{max}}$ is measured, $z_{\mathrm{dyn}}$ is the redshift one dynamical timescale ago from $z_{\mathrm{now}}$, and $z_{\mathrm{Mpeak}}$ is the redshift at $M_{\mathrm{Peak}}$. The ranks of $\Delta v_{\mathrm{max}}$ are essentially their probits according to the cumulative distribution function (CDF), which is assumed to be Gaussian. The probits are further mapped to the ranked percentiles of SFR by the UM DR1 model via a rank correlation parameter $r_{c}$ (see Equation 18 in B19) with some random Gaussian scatter. 

\begin{figure*}[h!]
    \centering
	\includegraphics[width=0.92\textwidth]{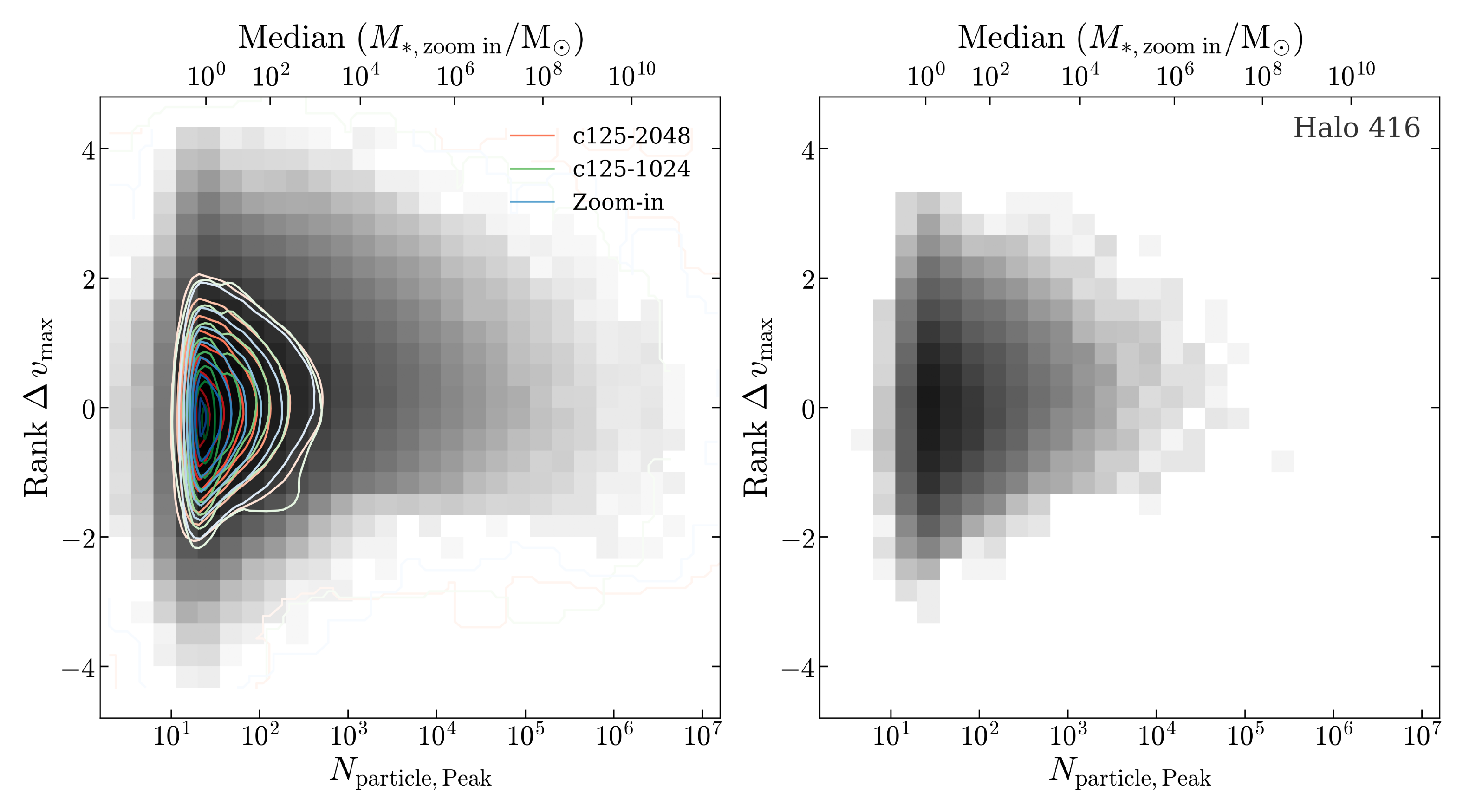}
    \caption{{\it Left panel:} Isodensity contours for the ranks of $\Delta v_{\mathrm{max}}$-$M_{\mathrm{Peak}}/m_{\mathrm{DM}}$ for (sub)halos in \textsc{c125-2048} (red), \textsc{c125-1024} (green), and joint MW resimulations (blue). The grey-scale color map shows the binned density of ranks versus peak halo particle number for (sub)halos in the joint zoom-in simulations. {\it Right panel:} Binned density map for ranks of $\Delta v_{\mathrm{max}}$ for (sub)halos in the zoom-in MW halo No.\ 416 (shown in Fig.~\ref{fig:2}. The vertical axes in both panels are in units of the standard deviation of the underlying CDF of $\Delta v_{\mathrm{max}}$. The top axes in both panels indicate the corresponding median stellar mass from the median SMHM relation in Fig.~\ref{fig:1} for the joint zoom-ins.}
    \label{fig:A1}
\end{figure*}

Shown in Fig.~\ref{fig:A1} are the rank distributions of $\Delta v_{\mathrm{max}}$ as a function of the peak halo particle number ($M_{\mathrm{Peak}}/m_{\mathrm{DM}}$) for the 45 joint zoom-ins (left panel) and the individual MW zoom-in halo No. 416 (right panel). The colored curves denote the isodensity contours of the rank distributions in the three resolutions, while the grey-scale color map shows the binned rank versus peak halo particle number for the zoom-ins (all subhalos in the left panel, and only those in Halo no. 416 on the right). We notice from the left panel that the contours for the subhalo ranks we obtain in the joint zoom-ins are close to those obtained in the two parent boxes. This is an interesting feature that could be useful when multiple zoom-ins are not available for ranking (sub)halos, and one could potentially use ranks obtained for (sub)halos with the same particle number from the parent box(es) as ranks assigned to zoom-in (sub)halos. However, this might only apply to low-mass subhalos ($M_{\mathrm{Peak}}\lesssim 10^{11}\mathrm{M_{\astrosun}}$) in MW zoom-ins due to the weak mass-dependence of halo concentration~(e.g., \citealt{2002ApJ...568...52W,2015ApJ...799..108D}) in this mass regime. We leave a quantitative investigation of this $\Delta v_{\mathrm{max}}$ ranking feature to future work.

Comparing the left and right panels, the most significant difference comparing the rank distributions of the joint and individual zoom-ins is that the joint zoom-ins' ranks have a much better coverage at the high ($\gtrsim 10^{5}$) and low ($\gtrsim 10$) particle number numbers, where the individual zoom-ins easily miss out on these rare objects due to finite sampling for high-mass objects or halo finder cut off for low-mass objects. In the range of $10^{1}\sim10^{3}$ particles where most (sub)halos dwell, the joint zoom-ins also sample the tails of the CDF more accurately ($\geqslant |4\sigma|$). In practice, if the tails are not sampled properly by applying \textsc{UniverseMachine} to an individual zoom-in, the code can break down due to the presence of rapidly growing ($>+7\sigma$) or heavily tidal-stripped ($<-7\sigma$) (sub)halos before fully evolving to $z=0$. Therefore, it is crucial to join multiple zoom-in realizations for obtaining unbiased ranks of the subhalos when applying empirical models like \textsc{UniverseMachine} to zoom-in simulations in general.

\section{Resolution cuts}
\label{sec:A2}

\begin{figure*}
    \centering
    \includegraphics[width=\textwidth]{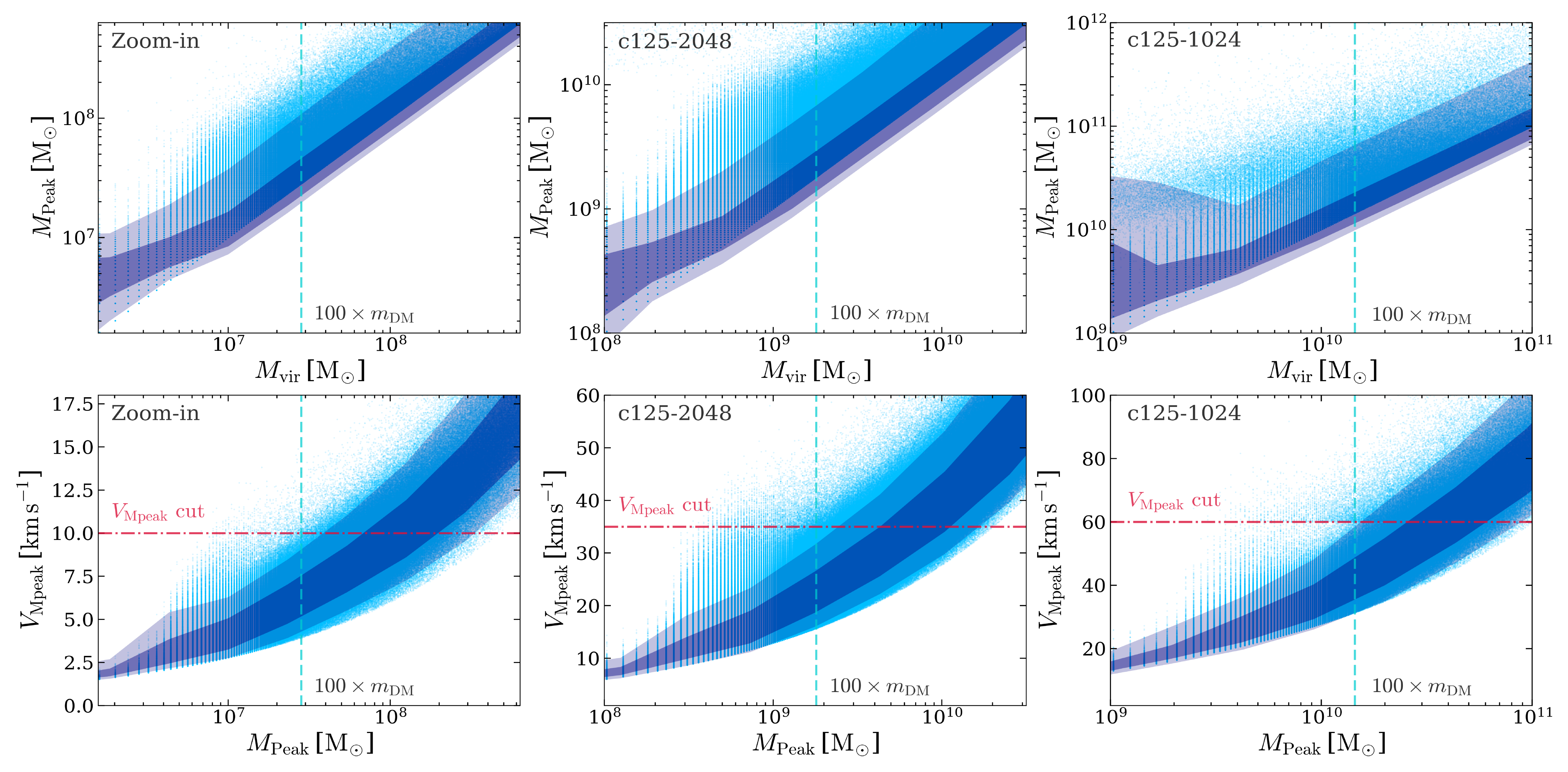}
        \caption{Peak halo mass $M_{\mathrm{Peak}}$ versus the maximum circular velocity at peak halo mass $V_{\mathrm{Mpeak}}$, and the virial mass $M_{\mathrm{vir}}$ at $z=0$ for all (sub)halos in \textsc{c125-1024}, \textsc{c125-2048}, and the joint MW zoom-ins at $z=0$. Dark and light shaded regions indicate the 68\% ($1\sigma$) and 95\% ($2\sigma$) intervals for the distributions. The vertical dashed line in each panel indicates $100\times$ the mass resolution, while the red dotted-dashed lines indicate the adopted $V_{\mathrm{Mpeak}}$ cut for well resolved halos in the three different resolutions. This selects most of the halos with more than 100 particles not only at their peak mass but also their present-day $M_{\mathrm{vir}}$ for all three resolutions.}
        \label{fig:A2}
\end{figure*}

Although our results are hardly affected by resolution effects on our galaxy and halo samples, we emphasize that one should be cautious when interpreting the results involving a significant proportion of (sub)halos near the resolution limit of the simulation. Numerical resolution mainly affect the density profiles of dark matter halos by decreasing the central region density and henceforth decreasing the $v_{\mathrm{max}}$. In Fig.~\ref{fig:A2} we show the $M_{\mathrm{vir}}$-$M_{\mathrm{Peak}}$ and $V_{\mathrm{Mpeak}}$-$M_{\mathrm{Peak}}$ relations for our three different resolution simulations. Scattered dots indicate individual (sub)halos, while the dark and light shaded regions indicate the $1\sigma$ and $2\sigma$ intervals of the corresponding distributions. 

Conventionally, halos with more than 100 particles at $z=0$ are thought to be well resolved. However, (sub)halos may experience mass loss due to tidal interactions in the cosmic environment, therefore one must also include (sub)halos that have well-resolved evolution histories but currently stripped below the traditional resolution limit. We propose that a reasonable resolution cut for our set of simulations is to select (sub)halos above the upper bound of the $2\sigma$ interval of $V_{\mathrm{Mpeak}}$ at the peak halo mass, which is 100 times that of the particle mass resolution. These values are indicated by the red dotted dashed lines in the lower panels of Fig.~\ref{fig:A2} (\textsc{c125-1024} $\sim 60$ km/s, \textsc{c125-2048} $\sim 35$ km/s, and zoom-ins $\sim 10$ km/s). This procedure deliberately selects subhalos with stellar mass ($\propto V_{\mathrm{Peak}}$) that up-scatters in the SMHM relation (which is acceptable for dark matter related properties). One could potentially also cut conservatively on $M_{\mathrm{Peak}}$ with $\gtrsim 200\times$ the particle mass resolution if one is particularly interested in (sub)halos with stellar masses near the $V_{\mathrm{Mpeak}}$ resolution limit. 

\end{document}